\documentclass{PoS}
\usepackage{mathtools}
\usepackage{graphicx, amssymb, amsfonts, amsmath, wrapfig}
\def\nn{\nonumber}

\def\amulohvp{a_\mu^\text{{\tiny LO-HVP}}}

\def\amuudlohvp{a_{\mu,{\rm ud}}^{\text{{\tiny LO-HVP}}}}
\def\amuslohvp{a_{\mu,{\rm s}}^{\text{{\tiny LO-HVP}}}}
\def\amuclohvp{a_{\mu,{\rm c}}^{\text{{\tiny LO-HVP}}}}
\def\amudisclohvp{a_{\mu,{\rm disc}}^{\text{{\tiny LO-HVP}}}}
\def\amuuddisclohvp{a_{\mu,{\rm ud/disc}}^{\text{{\tiny LO-HVP}}}}
\def\aelllohvp{a_\ell^\text{{\tiny LO-HVP}}}

\definecolor{indigo-dye}{rgb}{0.0, 0.25, 0.42} 
\definecolor{darkspringgreen}{rgb}{0.09, 0.45, 0.27} 

\title{
Review of Lattice QCD Studies of Hadronic Vacuum Polarization Contribution to Muon $g-2$
}
\ShortTitle{LQCD Studies for HVP \& Muon $g-2$}

\author{
\speaker{
Kohtaroh Miura
}\\ 
Helmholtz-Institut Mainz, Johannes Gutenberg-Universit\"{a}t Mainz, D-55099 Mainz, Germany\\
Kobayashi-Maskawa Institute for the Origin of Particles and the Universe, Nagoya University, Nagoya 464-8602, Japan\\
E-mail: \email{kohmiura@uni-mainz.de}
} 

\abstract{
Lattice QCD (LQCD) studies for the hadron vacuum polarization (HVP) and its contribution to
the muon anomalous magnetic moment (muon $g-2$) are reviewed.
There currently exists more than 3-$\sigma$ deviations in the muon $g-2$ between the BNL experiment
with 0.5 ppm precision and the Standard Model (SM) predictions,
where the latter relies on the QCD dispersion relation for the HVP.
The LQCD provides an independent crosscheck of
the dispersive approaches and important indications for assessing the SM prediction with measurements
at ongoing/forthcoming experiments at Fermilab/J-PARC (0.14/0.1 ppm precision).
The LQCD has made significant progress, in particular,
in the long distance and finite volume control,
continuum extrapolations, and QED and strong isospin breaking (SIB) corrections.
In the recently published papers, 
two LQCD estimates for the HVP muon $g-2$ are consistent with {\em No New Physics} while the other three are not.
The tension solely originates to the light-quark connected contributions
and indicates some under-estimated systematics in the large distance control.
The strange and charm connected contributions as well as the disconnected contributions are consistent
among all LQCD groups and determined precisely.
The total error is at a few percent level.
It is still premature by the LQCD to confirm or infirm the deviation between the experiments and the SM predictions.
If the LQCD is combined with the dispersive method, the HVP muon $g-2$ is predicted with $0.4\%$ uncertainty,
which is close upon the target precision required by the Fermilab/J-PARC experiments.
Continuous and considerable improvements are work in progress,
and there are good prospects that the target precision will get achieved within the next few years.
}

\FullConference{The 36th Annual International Symposium on Lattice Field Theory - LATTICE2018\\
		22-28 July, 2018\\
		Michigan State University, East Lansing, Michigan, USA.}

\begin{document}
\section{Introduction}
\begin{wrapfigure}{r}{0.3\textwidth}
\vspace{-4mm}
\begin{center}
\includegraphics[width=0.3\textwidth]{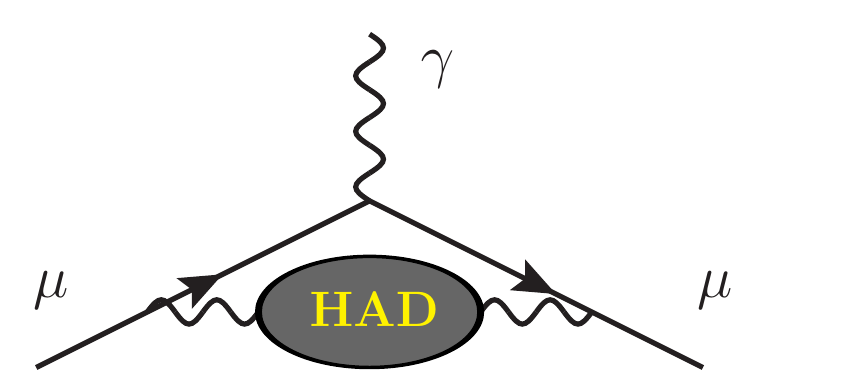}
\caption{The HVP contributions to muon $g-2$.}\label{fig:hvp}
\end{center}
\vspace{-4mm}
\end{wrapfigure}
Since the discovery of the Lamb shift,
the anomalous magnetic moment of charged leptons $(a_{\ell} = (g_{\ell} - 2)/2,\ g_{\ell} = \text{gyro-magnetic factor})$
have accompanied the development of quantum field theory (QFT).
For electrons ($a_{\ell = e}$), it is one of the most precisely measured and computed quantities in science with a total
uncertainty below 1 ppb. Theory and experiment agree, which is a great success of the QFT
and the standard model of particle physics (SM).
\footnote{
It is pointed out~\protect\cite{Meyer:2018til} that
the improved determination of the fine structure constant~\protect\cite{Parker:2018vye}
results in 2.4-$\sigma$ deviation: $a_e^{\rm exp} - a_e^{\rm SM} = (-87\pm 36)\times 10^{-14}$.}
The gyro-magnetic factor $g_{\ell}$ relates the lepton spin to its magnetic moment:
$\vec{M} = g_{\ell}(e_{\rm el}/(2m_{\ell}))\vec{s}$ where $e_{\rm el}$ and $m_{\ell}$ denote
the electric charge and lepton mass, respectively.
A finite $a_{\ell}$ shows a deviation from $g_{\ell} = 2$ predicted by the Dirac theory
and accounts for quantum loop (QFT) corrections with particles in the SM
and possibly physics beyond the standard model (BSM).
With respect to the BSM search, the {\em muon} anomalous magnetic moment ($a_{\ell = \mu}$)
is under active scrutiny both theoretically and experimentally;
there currently exists tension of more than 3-$\sigma$ deviations
between the experiment with 0.5 ppm precision~\cite{Bennett:2006fi} and the SM prediction
(quoted from~\cite{Aoyama:2012wk}, see also~\cite{Davier:2010nc,Hagiwara:2011af}):
\begin{align}
10^{10}\ a_{\mu}^{\text{expr.}} =  116 592 08.9 \pm 6.3\ (0.54\text{ppm})\ ,\quad
10^{10}\ a_{\mu}^{\text{SM}} = 116 591 84.0 \pm 5.9\ (0.51\text{ppm})\ .\label{eq:amu_sm_expr}
\end{align}
In fact, $a_{\mu}$ is generically much more sensitive to the massive BSM particles than $a_{e}$
since the BSM contributions are proportional to the lepton mass squared, 
{\em i.e.} 40000 times larger for the muon.

In the theory side, the largest source of uncertainty (over 79\%) in the total error of $a_{\mu}^{\text{SM}}$
comes from the non-perturbative estimate of the leading-order (LO)
hadron vacuum polarization (HVP, $\hat{\Pi}(Q^2)$) contributions to the anomaly ($\amulohvp$, Fig.~\ref{fig:hvp}).
In Eq.~(\ref{eq:amu_sm_expr}), the $\amulohvp$ in the SM prediction utilizes the second-subtracted QCD
dispersion relation for the HVP,
\begin{align}
\label{eq:disp}
\hat{\Pi}(Q^2)
= \int_0^{\infty}ds\ \frac{Q^2}{s(s+Q^2)}\frac{\mathrm{Im}\Pi(s)}{\pi}
= \int_0^{\infty}ds\ \frac{Q^2}{s(s+Q^2)}\frac{-R(s)}{12\pi^2}\ .
\end{align}
The HVP imaginary part $\mathrm{Im}\Pi(s)$ describes decay resonances
and the second equality is the optical theorem $\mathrm{Im}\Pi(s) = -R(s)/(12\pi)$ 
with $R(s) = \sigma[s;\ e^+e^-\to \text{hadrons}]$ $/(4\pi\alpha(s)^2/(3s))$, where $\alpha = e_{\rm el}^2/(4\pi)$.
This {\em dispersive method} provides the most precise prediction of $\amulohvp$ today.
However, the systematics in $R(s)$ is challenging to be controlled.
In fact, some tension in the cross section data of $\sigma[e^+e^- \to \pi^+\pi^-]$
is reported and under debate~\cite{Ablikim:2015orh}.
This gives a source of systematics in the dispersive estimate of the $\hat{\Pi}$ and $\amulohvp$.

Lattice QCD (LQCD) which does not rely on any experimental inputs
can provide an independent cross-check of the dispersive approach.
Since the pioneering work~\cite{Blum:2004cq}, the LQCD has made significant progress~\cite{Meyer:2018til}.
The LQCD precision will become competitive to the dispersive method in the coming years.
The ultimate goal of the LQCD is to provide the $\amulohvp$ in $\mathcal{O}(0.1)\%$ precision
which is required with respect to the ongoing/forthcoming experiments;
1) in Fermilab (FNAL-E989), 
a new measurement of $a_{\mu}$ aiming at 0.14 ppm uncertainty has been started~\cite{Holzbauer:2017ntd},
2) in J-PARC-E34, an experiment with a new technology using an ultra-cold muon beam
is planed~\cite{Otani:2015jra} and aims at 0.1 ppm precision with completely different systematics
from both BNL and Fermilab.
In addition, the MUonE (CERN) project~\cite{Abbiendi:2016xup} plans to measure
the QED running coupling constant $\alpha(s)$ precisely and provides the HVP with spacelike low momenta,
which can be combined with the HVP by the LQCD and allow the precise estimate of the $\amulohvp$~\cite{Marinkovic:2018PoS}.
Moreover, the LQCD HVP is applied to the precision science of the weak running coupling constant
at low momenta~\cite{Ce:2018ziv}, which is another place to assess the SM with experimental data
(e.g. PRISMA-Mainz~\cite{Becker:2018ggl}).

This proceedings reviews LQCD results for the HVP $\hat{\Pi}(Q^2)$
and its contribution to the anomaly $\aelllohvp$ (in particular, muon case $\ell = \mu$).
Comparing to the recent comprehensive review~\cite{Meyer:2018til},
we concentrate on the LO-HVP contribution to the anomaly
and include updates in addition to the published ones.

\section{Methodology}\label{sec:method}

Consider a scattering process of a charged lepton ($\ell^{-}$) from a photon ($A_{\nu}$).
The amplitude is expressed as
$
i\mathcal{M}(p,p^{\prime})(2\pi)^4\delta^{(4)}_{p,p^{\prime}}
= -ie_{\rm el}\bar{u}_{\ell}(p^{\prime}) \Gamma_{\nu}(p^{\prime}, p)u_{\ell}(p)A^{\nu}(p-p^{\prime}),
$
where $u_{\ell}$ represents a Dirac spinor of the lepton,
$(\hspace{-0.5mm}\not\hspace{-1mm} p - m_{\ell})u_{\ell}(p) = 0$,
and $\Gamma^{\nu}$ denotes an electromagnetic vertex operator.
Figure~\ref{fig:hvp} is an example for the vertex function with HVP.
In QCD/QED, which preserves the $CP$ symmetry, $\Gamma^{\mu}$ consists of only electric and magnetic terms,
$\Gamma^{\mu}(p^{\prime},p) = F_1^{\ell}(q^2)\gamma^{\mu} + F_2^{\ell}(q^2)(i\sigma^{\mu\nu}q_{\nu}/(2m_{\ell}))$,
where $q^{\mu} = (p^{\prime} - p)^{\mu}$ and $F_1^{\ell}$ ($F_2^{\ell}$) is known as the Dirac (Pauli) form factor.
The anomaly is then obtained from the form factor~\cite{Knecht:2003kc}:
$a_{\ell} = (g_{\ell} - 2)/2 = F_2^{\ell}(0)$ ({\em c.f.} $F_1^{\ell}(0) = 1$ in all order).
%
%
At tree level, $\Gamma^{\mu} = \gamma^{\mu}$ and $a_{\ell} = 0$.
At one-loop, the vertex function $\Gamma^{\mu}$ is the one without HVP blob in Fig.~\ref{fig:hvp},
giving the famous Schwinger's result:
$a_{\ell}^{1l} = (\alpha/\pi)\int_0^{\infty}d\hat{s}\ K(\hat{s}) = \alpha/(2\pi)$ with $\alpha = e_{\rm el}^2/(4\pi)$.
For the kernel $K(\hat{s})$, see the followings.
\footnote{In the one-loop calculation, the lepton mass $m_{\ell}$ is absorbed into
the integral variable $\hat{s} = Q^2/m_{\ell}^2$ and the Schwinger's result holds for arbitrary charged leptons: $e/\mu/\tau$.}

\subsection{Target Quantity}\label{subsec:target}
Our target is the leading-order (LO, $\sim \mathcal{O}(\alpha^2)$)
HVP contribution to the charged lepton (in particular, muon) $g-2$ shown in Fig.~\ref{fig:hvp},
and obtained by inserting a photon-irreducible HVP function ($\hat{\Pi}$) into the one-loop expression,
\begin{align}
& \aelllohvp
= \Bigl(\frac{\alpha}{\pi}\Bigr)^2 \int_0^{\infty}dQ^2 \frac{K(\hat{s})}{m_{\ell}^2} \hat{\Pi}(Q^2)\ ,\quad
\hat{\Pi}(Q^2) = 4\pi \bigl(\Pi(Q^2) - \Pi(0)\bigr)\ ,\label{eq:aelllohvp}\\
& K(\hat{s})
= \hat{s} (Z(\hat{s}))^2 \frac{1 - \hat{s}Z(\hat{s})}{1 + \hat{s}(Z(\hat{s}))^2}\ ,\quad
Z(\hat{s}) = -\frac{\hat{s} - \sqrt{\hat{s}^2 + 4\hat{s}}}{2\hat{s}}\ ,\quad
\hat{s} = \frac{Q^2}{m_{\ell}^2}\ ,\label{eq:ker}
\end{align}
where $\Pi(Q^2)$ is a scalar part in the electromagnetic vector current ($j_{\mu}$) correlator
with up, down, strange, and charm quarks ($\psi_{f=u,d,s,c}(x)$),
\begin{align}
& \Pi_{\mu\nu}(Q)
= (\delta_{\mu\nu} Q^2 - Q_{\mu}Q_{\nu})\Pi(Q^2)
= \int d^4x\ e^{iQ\cdot x} G_{\mu\nu}(x)\ ,\quad
G_{\mu\nu}(x)
= \big\langle \mathrm{Re} [j_{\mu}(x)j_{\nu}(0)]\big\rangle\ ,\label{eq:Pimn}\\
& j_{\mu}(x) =
\sum_{f = u,d,s,c} q_f j_{\mu}^f(x) =
\sum_{f}q_f\ (\bar{\psi}_f\gamma_{\mu}\psi_f)(x)\ ,\quad
(q_{u,c}, q_{d,s}) = (2/3, -1/3)\ .\label{eq:jmu}
\end{align}
In the muon case $a^{\text{{\tiny LO-HVP}}}_{\ell = \mu}$,
the required precision is $\mathcal{O}(0.1) \%$
which amounts to $0.1$ ppm
(target precision of the experiments~\cite{Holzbauer:2017ntd,Otani:2015jra}) in the total $a_{\mu}$.
The bottom- and top-quark effects are negligible for that precision
(the bottom quark contribution is estimated to be below 0.05\% in $\amulohvp$~\cite{Chakraborty:2016mwy}).
In contrast to the one-loop expression mentioned above,
the integrand in Eq.~(\ref{eq:aelllohvp}) is not a function of the single variable $\hat{s}$;
the lepton mass $m_{\ell}$ sets a typical scale via $K(\hat{s})/m_{\ell}^2$
and the integrand has a peak around $Q^2 = (m_{\ell}/2)^2$.
Once the HVP $\hat{\Pi}(Q^2)$ is calculated,
the anomaly for arbitrary leptons $a_{\ell = e/\mu/\tau}$ could get available
as long as systematics in $\hat{\Pi}(Q^2)$ near $Q^2 = (m_{\ell}/2)^2$ are controlled.

We shall define ``the LO-HVP contribution - $\aelllohvp$'' more precisely;
\textcircled{\scriptsize 1} as shown in Fig.~\ref{fig:hvp},
only two internal photons connected with the HVP blob attach the external lepton lines, and
\textcircled{\scriptsize 2} only one HVP blob which is irreducible with respect to a photon line cutting is inserted.
Two photons in \textcircled{\scriptsize 1}  give rise to
$\mathcal{O}(\alpha^2)$ - the prefactor in Eq.~(\ref{eq:aelllohvp}).
The condition \textcircled{\scriptsize 2} excludes most of diagrams with additional $\alpha$
but still allows $\alpha$ associated with photons inside the single-irreducible HVP blob $\hat{\Pi}$.
We shall call them as {\em QED corrections in LO-HVP}.
From now on, our target quantity $\aelllohvp$ is considered
to include such QED corrections $\mathcal{O}(\alpha^{n\geq 3})$ in addition to the pure-QCD effects $\mathcal{O}(\alpha^2)$.
For the target precision, the leading QED correction must be considered:
$\mathcal{O}(\alpha^3)/\mathcal{O}(\alpha^2)\sim \alpha\sim 1\% > 0.1\%$.
In compering/combining the LQCD to the dispersive method (\ref{eq:disp}),
the QED corrections must be taken account in the LQCD side since
it is impossible to extract the pure QCD HVP alone in the dispersive method.
The other $\mathcal{O}(\alpha^3)$ diagrams,
with three internal photons attaching the leptons, two HVP blobs, or one HVP blob and one lepton loop independent of the HVP blob
are considered as next-to-leading-order (NLO) contributions, which is investigated in Ref.~\cite{Chakraborty:2018iyb}
and not studied in this proceedings.

The simulation output in LQCD is the correlator $G_{\mu\nu}$ in Eq.~(\ref{eq:Pimn}),
which is composed of the connected and disconnected correlators,
\begin{align}
\label{eq:Gt}
& G(t) = (a^3/3)\sum_{i = 1}^{3} \sum_{\vec{x}} G_{ii}(\vec{x},t) = (\sum_fC^f + D)(t)\ ,\\
& C^f(t) = -(a^3/3)\sum_{i = 1}^{3} q_f^2\sum_{\vec{x}}
\langle \mathrm{Re}\mathrm{Tr}[\gamma_iS^f(x,0)^{\dagger}\gamma_iS^f(x,0)]\rangle\ ,\\
& D(t) = (a^3/3)\sum_{i = 1}^{3}\sum_{f,f^{\prime}}q_fq_{f^{\prime}}\sum_{\vec{x}}
\langle \mathrm{Re}\{\mathrm{Tr}[\gamma_iS^f(x,x)]\mathrm{Tr}[\gamma_iS^{f^{\prime}}(y,y)]\}\rangle|_{y = 0}\ .
\end{align}
Here, $S^f(x,y)$ is a quark propagator ($\{\mathcal{D}^{f}(x,y)\}^{-1}$) with flavor $f$.
For obtaining $S^f(x,y)$, it is too expensive to solve the Dirac equation with a source position at every lattice points.
In practice, one usually reduces the cost by introducing stochastic sources scattered in the 3- or 4-volume of the lattice
and solves the stochastic Dirac equation $\mathcal{D}^{f}(y,z)\phi^{(r)}(z) = \eta^{(r)}(y)$
where $r\in [1, N_r]$ and $\eta^{(r)}(y)$ is a stochastic noise vector satisfying the condition:
$\lim_{N_r\to\infty}N_r^{-1}\sum_{r = 1}^{N_r}\eta^{(r)}(x)\eta^{(r)}(y)^{\dagger} = \delta_{xy}$.
The number of solving the Dirac equation decreases to $N_r$ but the additional noises from $\eta^{(r)}$ are introduced.
Thus, the cost problem is altered into the noise problem.
For $D(t)$~\cite{Francis:2014hoa,Blum:2015you,Blum:2018mom,Yamamoto:2018cqm,Borsanyi:2017zdw,Borsanyi:2016lpl},
one needs to estimate all-to-all correlators of the quark-loop
$l_{\mu}^f(t) = a^3\sum_{\vec{x}}\mathrm{Tr}[\gamma_{\mu}S^f(x,x)]$
and the stochastic estimate
$l_{\mu}^f(t) = a^3\sum_{\vec{x}}N_r^{-1}\sum_{r=1}^{N_r}\mathrm{Tr}[\eta^{(r)}(x)^{\dagger}\gamma_{\mu}\phi^{(r)}(x)]$
decreases the computational time by a factor of $N_r / \text{4-Volume} \sim \mathcal{O}(10^{3 - 4} / 10^7)$
but the noise problem gets serious.
To suppress the noise, the stochastic method is usually combined
with the all-mode-averaging (AMA) technique~\cite{Blum:2012uh}
and/or the hierarchical technique~\cite{Stathopoulos:2013aci}.
In the isospin limit, light and strange quark contributions is simplified as
$D(t) = (1/9)\sum_{\vec{x}}(\sum_{i=1}^{3}/3)
\big\langle\mathrm{Re}[j_{i}^{\rm disc}(\vec{x},t)j_{i}^{\rm disc}(0)]\big\rangle$
where $j_{\mu}^{\mathrm{disc}}(x) = \mathrm{Tr}[\gamma_{\mu}S^{ud}(x,x)] - \mathrm{Tr}[\gamma_{\mu} S^s(x,x)]$.
Using common noise vectors $\eta^{(r)}$ in calculating the first and second terms in $j_{\mu}^{\rm disc}$,
the stochastic noises in $D(t)$ largely cancel out between them~\cite{Francis:2014hoa}.
\footnote{
The charm quark contribution to $D(t)$ may be evaluated separately
by hopping-parameter expansions~\protect\cite{Borsanyi:2017zdw,Borsanyi:2016lpl}.}

We shall derive formulas~\cite{Bernecker:2011gh} relating the LQCD output $G(t)$
to our target quantities, $\amulohvp$ and $\hat{\Pi}$.
We work in the continuous spacetime and shows in later the formulas applies to the lattice system.
First, we use the property of
$\Pi_{\mu\nu}(\omega\to 0) = \int d^4x G_{\mu\nu}(x) = \int d^4x \partial_{\rho}(x_{\mu}G_{\rho\nu}(x)) = 0$,
where the second equality results from the vector current conservation $\partial_{\rho}j_{\rho}(x) = 0$.
Second, specify the momentum as $(Q_{\mu}) = (\omega,Q_{i = 1,2,3}) = (\omega,\vec{0})$ in Eq.~(\ref{eq:Pimn})
which leads to $\Pi(\omega^2) = \sum_{i}\Pi_{ii}(\omega)/(3\omega^2)$.
And finally, note the fact that $\Pi(\omega^2)$ is an even function of $\omega$.
From these properties,
the Fourier transformation in Eq.~(\ref{eq:Pimn}) reduces into
$\Pi(\omega) = \int d\omega(\cos[\omega t] - 1) \mathcal{G}(t)/\omega^2$ with
$\mathcal{G}(t)$ being the continuum counterpart of $G(t)$ in Eq.~(\ref{eq:Gt}).
\footnote{
If one investigates HVP with an axial vector current ($j_{\rho,{\rm axial}}^{a}$),
the current conservation does not hold due to
the chiral symmetry breaking: $\partial_{\rho}j_{\rho}^{a,{\rm axial}}(x) = (m_u + m_d)(\bar{\psi}(\tau^a/2)\gamma_5\psi)(x)$.
One finds $\Pi_{\mu\nu}^{\rm axial}(Q\to 0) \neq 0$ unlikely to the vector case.
As a result, the Fourier transformation reduces into the $\cos[\omega t]$ term alone without ``$-1$''.}
Using $\Pi(\omega\to 0) = -\int dt (t^2/2)\mathcal{G}(t)$, we obtain~\cite{Bernecker:2011gh}
\begin{align}
\label{eq:tmr}
\hat{\Pi}(\omega^2) = \int_0^{\infty} dt\ t^2\bigl(1 - {\rm sinc}^2[\omega t/2]\bigr) \mathcal{G}(t)\ ,
\end{align}
where ${\rm sinc}[x] = x^{-1}\sin x$, and
\begin{align}
\label{eq:tmr_a}
\aelllohvp = \Bigl(\frac{\alpha}{\pi}\Bigr)^2\int_0^{\infty}dt\ W(t,m_{\ell})\mathcal{G}(t)\ ,\
W(t,m_{\ell})
= \int_0^{\infty}d\omega^2 \frac{K\bigl(\frac{\omega^2}{m_{\ell}^2}\bigr)}{m_{\ell}^2}\ 
t^2\Bigl(1 - {\rm sinc}^2\Bigl[\frac{\omega t}{2}\Bigr]\Bigr)\ .
\end{align}
The integrands in Eqs.~(\ref{eq:tmr}) and (\ref{eq:tmr_a}) are regular
at $t \to 0$, $t\to \infty$, $\omega\to 0$, and/or $\omega\to\infty$.

The above equations are the expressions in the continuum and infinite volume limits.
In LQCD with a finite volume (FV, $L^3\times T$) and lattice spacing ($a$),
the Ward-Takahashi Identity (WTI) for lattice symmetries,
$\sum_{\mu}\hat{Q}_{\mu}\Pi_{\mu\nu}(Q) = 0$ with $\hat{Q}_{\mu} = (2/a)\sin[(a/2)(2\pi n/L_{\mu})]$,
does not exclude longitudinal components of $\Pi_{\mu\nu}$,
a part of which becomes a large FV effect at infra-red limits.
\footnote{Besides FV effects, the WTI with a finite lattice spacing allows
terms proportional to $d^n\Pi/dQ^{2n}$ in Eq.~(\protect\ref{eq:Pimn}).
However, they have dimension $(-2n)$ and vanishes at least quadratically ($n\geq 1$) in the continuum limit.
In considering $\amulohvp$, the HVP with low momenta dominantly contributes,
and the derivative terms would be negligible~\protect\cite{Aubin:2015rzx}.}
Replacing $\Pi_{\mu\nu}(Q^2)$ with
$\bar{\Pi}_{\mu\nu}(Q^2) = \Pi_{\mu\nu}(Q^2) - \Pi_{\mu\nu}(0)$ in Eq.~(\ref{eq:Pimn}),
the derivation of Eqs.~(\ref{eq:tmr}) and (\ref{eq:tmr_a}) can be repeated in the LQCD case.

We shall consider the Taylor expansion of HVP:
$\hat{\Pi}(\omega^2)=\sum_{n\geq 1}^{\infty}\Pi_n\omega^{2n}$.
Using Eq.~(\ref{eq:tmr}) in the LQCD case ($t = \hat{t}a$, $\hat{t}\in \mathbb{Z}$),
the coefficients $\Pi_n$ are evaluated from the LQCD output $G(t)$,
\begin{align}
\label{eq:Pin}
\Pi_n
= \frac{1}{n!}\frac{d^n\hat{\Pi}}{d\omega^{2n}}\Big|_{\omega^2\to 0}
= \sum_{t}a \frac{(-t^2)^{n+1}}{(2n + 2)!} G(t)\ .
\end{align}
The integrand in the right hand side has a peak,
which is shifted to the larger $t$ direction with increasing $n$
because of the factor $(t^{2})^{n+1}$.
For example, the peaks with $n = 1$ and $2$ locate at $t \sim 1$ fm and $1.5$ fm, respectively.
Each time-moment $\Pi_{n = 1,2,\cdots}$ is responsible for the scale corresponding to the peak,
and allows a scale-by-scale comparison among various groups ({\em c.f.} Table~\ref{tab:tm_fbf}).
Moreover, the slope $\Pi_1$ is an important LQCD input for the combined estimate of the $\amulohvp$
with the LQCD, experimental data, and QCD sum rule~\cite{Dominguez:2017yga}.
As shown in the followings, $\Pi_n$ provides a key ingredient in several approximants in the IR region.

\subsection{Approximants}\label{subsec:approximant}
In the muon case, the typical scale $Q^2 = (m_{\ell = \mu}/2)^2$ in the integral (\ref{eq:aelllohvp})
is less than half of a pion mass squared ($m_{\pi}^2 / 2$).
In LQCD with finite volume ($L^3\times T$), it is hopeless to get enough data points around that momentum region
({\em c.f.} $Lm_{\pi}\lesssim 4 - 6$).
One needs some approximants for $\hat{\Pi}(Q^2\sim (m_{\mu}/2)^2)$.
To minimize associated systematics, model independent approximants
(e.g. unlikely to the vector-meson dominance) are preferable and reviewed in the followings.

\begin{wrapfigure}{r}{0.52\textwidth}
\vspace{-6mm}
\hspace{-5mm}
\begin{center}
\includegraphics[width=0.52\textwidth]{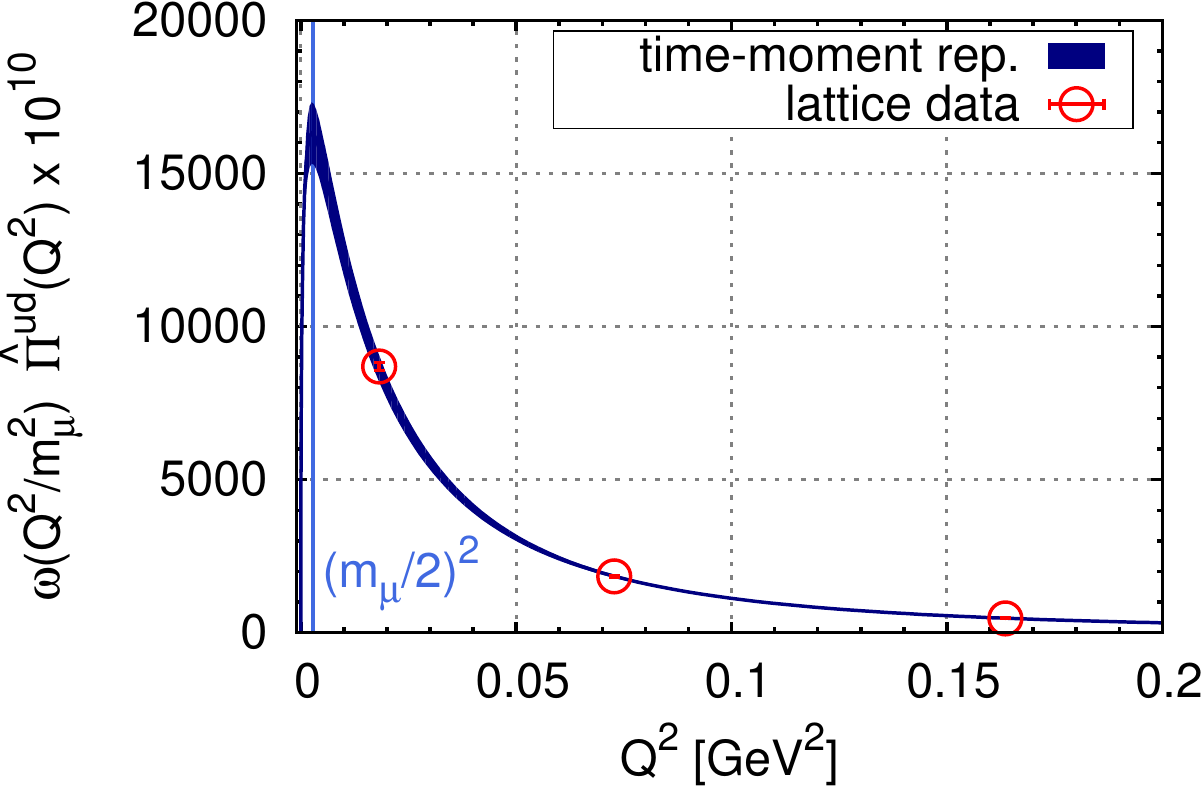}
\caption{
An example of TMR using BMW ensemble~\protect\cite{Borsanyi:2017zdw,Borsanyi:2016lpl}
with finest lattice spacing $a = 0.064$ fm.
The integral over $Q^2$ gives $\amulohvp$.
}\label{fig:tmr}
\end{center}
\vspace{-5mm}
\end{wrapfigure}

\paragraph{Time-Momentum Representation~\cite{Bernecker:2011gh}}
In the LQCD estimate of $\amulohvp$,
introducing the strict lattice momenta is not mandatory;
in Eq.~(\ref{eq:tmr}), replace the continuous $t$ with the lattice one $\hat{t}a$ ($\hat{t} \in \mathbb{Z}$)
but keep {\em continuous-valued} $\omega$,
$\hat{\Pi}^{\rm TMR}(\omega^2) = 
a^3\sum_{\hat{t}=0}^{T/2}\hat{t}^2 \bigl(1 - {\rm sinc}^2[\omega \hat{t}a/2]\bigr) G(\hat{t})$,
which is called the time-momentum representation (TMR)~\cite{Bernecker:2011gh}, and one of the popular choices today.
The approximants converges to the correct continuum limit.
Example of the integrand in Eq.~(\ref{eq:aelllohvp}) with $\hat{\Pi}^{\rm TMR}(\omega^2)$ is shown in Fig.~\ref{fig:tmr}.
In LQCD, we need an IR-cut in $\sum_{\hat{t}}$ and a UV-cut $\omega_{\rm cut} \lesssim \pi / (2a)$
in the integral in Eq.~(\ref{eq:tmr_a}) to control the associated systematics
(see, for example, the supplemental material of Ref.~\cite{Borsanyi:2017zdw}).

\paragraph{Pad\'{e} Approximants~\cite{Aubin:2012me}}
To investigate the deep IR region, $Q^2 \sim (m_{\mu}/2)^2$,
it is natural to utilize the zero-momentum Taylor coefficients (\ref{eq:Pin}).
We shall consider the $[M,N]$ Pad\'{e} approximants,
$\hat{\Pi}(Q^2;M,N) = (\sum_{m\geq 1}^{M} A_m Q^{2m})/(1 + \sum_{n\geq 1}^{N} B_n Q^{2n})$
($A_0 = 0$ by definition, $\hat{\Pi}(0) = 0$).
The coefficients $(A_m,B_n)$ are then constructed from the Taylor coefficients $\Pi_{n}$ (\ref{eq:Pin})
(the generalization $\omega^2\to Q^2$ is trivial).
Consider the first two approximants:
$\hat{\Pi}(Q^2;1,0) = Q^2\Pi_1$ and
$\hat{\Pi}(Q^2;1,1) = \Pi_1 Q^2 / (1 - Q^2\Pi_2/\Pi_1)$.
If the $\hat{\Pi}(Q^2;1,0)$ is adopted for whole $Q^2 = \omega^2$ in Eq.~(\ref{eq:aelllohvp}),
the integral gives the upper bound on the true value of $\amulohvp$.
Else if the $\hat{\Pi}(Q^2;1,1)$ is adopted,
the integral (\ref{eq:aelllohvp}) accounts for more than 98\% of the integral with full $\hat{\Pi}(Q^2)$.
The HVP $\hat{\Pi}(Q^2)$ satisfies the dispersion relation (\ref{eq:disp}),
which is seen as a Stielthes integral~\cite{Aubin:2012me} with positivity of the integrand $\mathrm{Im}\Pi$.
This property guarantees that $\hat{\Pi}(Q^2)$ is approximated by Pad\'{e} approximants with
{\em a finite convergence radius}.
Usually, $[2,1]$ or $[1,1]$ Pad\'{e} approximant is used in the {\em hybrid method};
use Pad\'{e} approximants for IR region ($Q^2 < Q_{\rm IR}^2\sim 0.2\ \text{GeV}^2$) in Eq.~(\ref{eq:Pimn}),
adopt a trapezoid (or similar) integral for intermediate momenta
$Q^2\in [Q^2_{\rm IR}, Q^2_{\rm UV}]$ with $Q^2_{\rm UV}\sim 4\ \text{GeV}^2$,
and evaluate higher momentum contributions with perturbative QCD.
The systematics is estimated by varying the cuts $Q^2_{\rm IR/UV}$.
One could also use Pad\'{e} approximants for whole $Q^2$ region by assuming that the true integral
exists in between the integrals obtained with $[N,N]$- and $[N,N-1]$-th approximants, whose difference is used
to estimate the systematic error~\cite{Chakraborty:2016mwy}.

\paragraph{Others}
Two other approximants
- the Mellin-Barnes (MB) approximants and the Lorentz-covariant coordinate-space (CCS) representation -
have been proposed.
The MB approximants are based on the fact that the spectrum $\mathrm{Im}\Pi(s)/\pi$ in QCD
is positive and approaches a constant as $s\to\infty$, and thereby, reproduce a relevant asymptotic behavior at large $s$
in contrast to some classes of Pad\'{e} approximants (e.g.~$[N,N-1]$-Pad\'{e}).
See, Ref.~\cite{Charles:2017snx} for details.
The MB approximants can be constructed by using the moments (\ref{eq:Pin}) calculated by LQCD
for which one must confirm that the convergence condition (Sec.~III.B in Ref.~\cite{Charles:2017snx}) is satisfied.
The MB has been adopted in the phenomenological estimate of $\amulohvp$~\cite{Benayoun:2016krn}.
In LQCD, the MB is calculated in Ref.~\cite{Giusti:2018mdh} and utilized to evaluate the NLO $\mathcal{O}(\alpha^3)$ HVP
contributions~\cite{Chakraborty:2018iyb}.
Another approximants, the CCS representation, introduces IR-cuts
in the radial coordinate for the hyper-sphere, $r = (\sum_{\mu = 0}^{3}x_{\mu})^{1/2}$,
and respects the Lorentz covariant expression. See Ref.~\cite{Meyer:2017hjv} for details.
In contrast to the TMR explained above, the IR-cut in the CSS excludes noises at large distance
in all of ${t,x,y,z}$ directions. Therefore, the CSS could achieve a better noise control, particularly for
the quark disconnected contributions, which are known to be notoriously noisy.
So far, the CCS has been applied to the low energy running of the weak mixing angle~\cite{Ce:2018ziv}.

\section{Challenges}\label{sec:challenge}
In the recent years, the LQCD has made significant progress,
particularly in the long distance and finite volume control,
continuum extrapolations, and QED and strong isospin breaking (SIB) corrections,
which will be reviewed in the followings.

\subsection{Long distance control}\label{subsec:long_distance}
As seen in Eq.~(\ref{eq:aelllohvp}), the typical scale $(m_{\mu}/(2\hbar c))^{-1}\sim 4$ fm
emerges from the kernel $K(\hat{s})/m^2_{\ell}$ in the calculation of $\amulohvp$.
The light connected and disconnected correlators ($C^{f = ud}(t), D(t)$) in Eq.~(\ref{eq:Gt})
includes 2-pion (isospin-one) contributions and some other light modes,
and sizable contributions to $\amulohvp$ remain at large distance of $3-4$ fm.
Even using many stochastic sources,
$C^{ud}(t)$ and $D(t)$ get strongly attenuated at $t \gtrsim 3$ fm and need further elaborations
in the precision ($\mathcal{O}(0.1)$ \%) science of the $\amulohvp$.
To this end, one approximates the original LQCD data of $C^{ud}(t)$ and $D(t)$ at large distance ($t\geq t_*$)
with modeling or reconstructing them so that the long distance properties are much better controlled than
those in the original correlators.
From a number of ideas proposed so far~\cite{Meyer:2018til},
we namely focus on the method using the time-like pion form factor
proposed by the Mainz group~\cite{DellaMorte:2017dyu}.

We shall consider the isospin decomposition of the correlator (\ref{eq:Gt}) with finite spatial extension $L$,
$G(t,L) = G^{I=1}(t,L) + G^{I=0}(t,L)$,
and investigate the isospin-one part $G^{I=1}(t,L)$ at large distance.
In the spectral representation, it is expressed as
\begin{align}
G^{I=1}(t,L) = \sum_{n=1} |A_n|^2 e^{-\omega_n t}\ ,\quad
\omega^2_n = 4(m_{\pi}^2 + k_n^2)\ .\label{eq:GI1}
\end{align}
We investigate the amplitudes $|A_n|$ and the energy levels $\omega_n$
via the L\"{u}scher's formula~\cite{Luscher:1991cf} in elastic cases and
Meyer's Formula~\cite{Meyer:2011um},
\begin{align}
& \delta_{l = 1}(k_n) + \phi\Bigl(\frac{k_nL}{2\pi}\Bigr) = n\pi\ ,\quad
\phi(z) = \frac{-z\pi^{3/2}}{(1/\sqrt{4\pi})\sum_{n\in\mathbb{Z}^3}(n^2 - z^2)^{-1}}\ ,\label{eq:luscher}\\
& |F_{\pi}(\omega_n)|^2 = \frac{3\pi\omega_n^2}{2k_n^5}
\bigl(\Delta \delta_1(k_n) + \Delta \phi(q_n)\bigr)|A_n|^2\ ,\quad
\Delta f(x) = \frac{df(x)}{d\log x}\ ,\label{eq:meyer}
\end{align}
as well as the Gounaris-Sakurai (GS) parameterization (c.f. Appendix of Ref.~\cite{Francis:2013qna}),
\begin{align}
\begin{cases}
\frac{k^3}{\omega}\mathrm{cot}\delta_1(k) =
k^2h(\omega) - k_{\rho}^2h(m_{\rho}) + (k^2 - k^2_{\rho})b\ ,\
b = \frac{-2}{m_{\rho}}
\Bigl(\frac{2k^3_{\rho}}{m_{\rho}\Gamma_{\rho}} + \frac{m_{\rho}h(m_{\rho})}{2} + k^2_{\rho}h^{\prime}(m_{\rho})\Bigr)\ ,\\
F_{\pi}(\omega)\sim F_{\pi}^{\rm GS}(\omega)
= \frac{-1}{(k^3/\omega)(\mathrm{cot}\delta_1(k) - i)}
\Bigl(
\frac{m^2_{\pi}}{\pi} + k^2_{\rho}h(m_{\rho}) + \frac{m^2_{\rho}b}{4}
\Bigr)\ ,\label{eq:gs}
\end{cases}
\end{align}
where $h(\omega) = (2k/(\pi\omega))\log[(\omega + 2k)/(2m_{\pi})]$,
$\omega^2 = 4(m^2_{\pi} + k^2)$, $k^2_{\rho} = (m^2_{\rho}/4 - m^2_{\pi})$,
and $\Gamma_{\rho}$ represents a rho-meson decay width.
For a given meson mass spectra $\{m_{\pi},m_{\rho}\}$,
the GS parameterization (\ref{eq:gs}) allows us to express the p-wave phase shift $\delta_1(k_n)$
and the time-like pion form factor $F_{\pi}(\omega_n(k_n))$
as a function of the decay width $\Gamma_{\rho}$ and lattice momenta $k_n$.
They are substituted into
the L\"{u}scher's formula (\ref{eq:luscher}) and Meyer's formula (\ref{eq:meyer})
to express $k_n$ and the amplitude $A_n$ as a function of $\Gamma_{\rho}$.
Then, the isospin-one correlator (\ref{eq:GI1}) solely depends on the single unknown parameter $\Gamma_{\rho}$
and fitted to the original LQCD data of $G^{I = 1}(t,L)$ to determine the $\Gamma_{\rho}$,
and equivalently, $(k_n, A_n)$.
The correlator $G^{I = 1}(t, L)$ in Eq.~(\ref{eq:GI1}) is now reconstructed with the obtained $(\omega_n(k_n), A_n)$.

\begin{wrapfigure}{r}{0.55\textwidth}
\vspace{-2mm}
\hspace{-4mm}
\begin{center}
\includegraphics[width=0.55\textwidth]{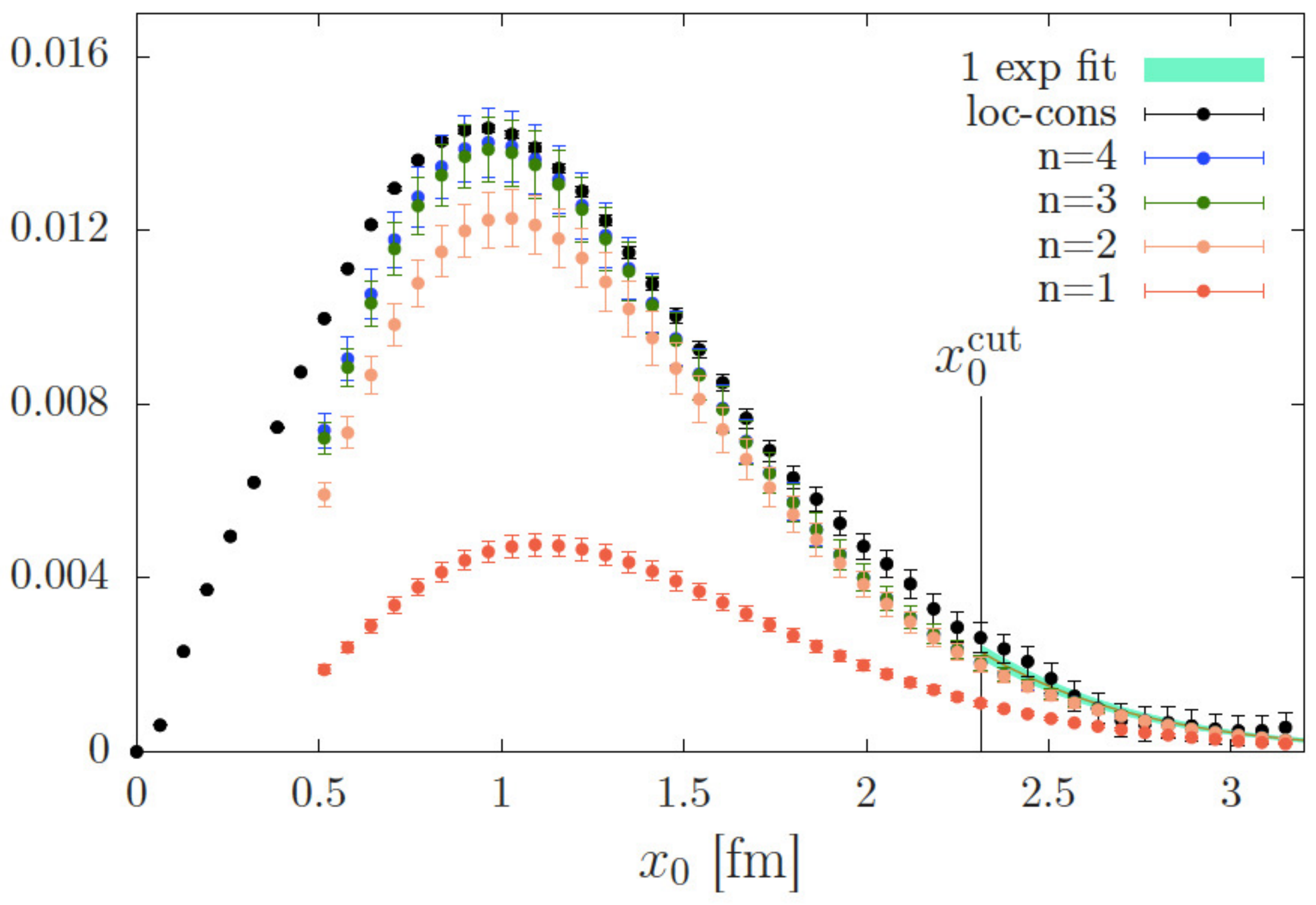}
\caption{
Reconstructed-accumulated correlators
$G^{I=1}_n(t,L) = \sum_{j=1}^{n} |A_j|^2 e^{-\omega_j t}$
multiplied by the weight function ($x_0 = t$), quoted from Ref.~\cite{Gerardin:2018sin}.
}\label{fig:lat2019_mainz_Gn}
\end{center}
\vspace{-4mm}
\end{wrapfigure}
With the obtained $(\omega_j, A_j)$,
investigate the accumulated correlator~\cite{Gerardin:2018sin}
$G^{I=1}_n(t,L) = \sum_{j=1}^{n} |A_j|^2 \exp[-\omega_j t]$.
Figure~\ref{fig:lat2019_mainz_Gn} displays
$\tilde{W}(t,m_{\ell})G^{I=1}_n(t,L)$ where $\tilde{W} \propto W$ in Eq.~(\ref{eq:tmr_a}).
The lightest mode $n = 1$ becomes dominant at quite large distance around $t\sim 3$ fm,
where the noise control of the original correlator is still challenging.
In turn, the sum of the lightest three modes ($j = 1 - 3, n = 3$)
gives a good approximation of the original correlator around the peak ($\sim 1$ fm) and larger distance.
For $t > x_0^{\rm cut}$ in the figure,
the accumulated correlators (e.g. $G^{I=1}_{n = 4}(t,L)$) have much smaller uncertainty and
give better estimates than the original one. The $\amuudlohvp$ with the $G^{I=1}_{n = 4}(t,L)$ used at large distance
becomes reliable and possesses a smaller statistical error. The systematic error is estimated by varying the threshold $x_0$.

In principle, the time-like pion form factor $F_{\pi}$ and the phase shift $\delta_1(\omega)$
can be directly investigated in LQCD without recourse to the GS representation.
In practice, however, one must study the energy spectra $\omega_n$ and the amplitudes $A_n$
by taking account of the pion scattering effects, which is challenging in finite volume Euclidean spacetime.
In Ref.~\cite{Erben:2017hvr}, $\omega_n$ and $A_n$ with rho meson resonance effects are extracted
by solving a generalized eigen value problem for the pion-rho meson correlator matrices created with a distillation technique.
Combined with the L\"{u}scher (\ref{eq:luscher}) and the Meyer (\ref{eq:meyer}),
one can determine the $(F_{\pi}, \delta_1)(\omega)$ based on the LQCD in the self-contained way.
The reconstructed correlators in this method are recently studied by RBC/UKQCD collaboration~\cite{Meyer:2018PoS}.
Further advanced analyses are work in progress. Consider the Omn\`{e}s formula:
$F_{\pi}(\omega) = \exp[\omega^2 P_{n-1}(\omega^2) + (\omega^{2n}/\pi)Q_n(\omega^2)]$
with $Q_n(\omega^2) = \int_{4m_{\pi}^2}^{\infty}ds\delta_1(s)/\{s^{n}(s - \omega^2 -i\epsilon)\}$.
For a selected subtraction level $n$, the Polynomial $P_{n-1}$ is determined by fitting
the formula to the LQCD data of $(F_{\pi},\delta_1)(\omega)$.
This method does not rely on the GS parameterization~\cite{Gerardin:2018sin}.
The long distance behavior of $C^{ud}(t)$ and $D(t)$
is related to FV effects, $\Delta\amuudlohvp(L_1,L_2) = \amuudlohvp(L_2) - \amuudlohvp(L_1)$
with $L_1 \ll L_2$, which are summarized in Table~\ref{tab:fv}.
As argued in~\cite{Aubin:2015rzx},
the FV effects at sufficiently large distance would be governed by pions,
in particular the $I = 1$ contributions (2-pion exchange with minimal lattice momentum $2\pi / L$),
which can be computed in the chiral perturbation theory (2-pion-XPT),
which results in $\sim 2\%$ correction to the total $\amulohvp$~\cite{Borsanyi:2017zdw}.
The FV effects taking account of more than the lowest 2-pion modes can be performed
by using the time-like pion form factor $F_{\pi}$;
the {\em infinite volume} isospin-one correlator at large distance is expressed as,
$G^{I=1}(t,L\to\infty) = (1/48\pi^2)\int_{2m_{\pi}}^{\infty}d\omega\
\omega^2\bigl(1 - 4m_{\pi}^2/\omega^2\bigr)^{3/2}|F_{\pi}(\omega)|^2 e^{-\omega |t|}$,
and comparing the $\amulohvp$ obtained by $G^{I=1}_n(t,L)$ and $G^{I=1}_n(t,\infty)$
gives FV estimates including excited modes ($n\geq 2$).
We call this method as Gounaris-Sakurai-L\"{u}scher-Meyer (GSLM) method,
which was proposed in the recent publication by Mainz group~\cite{DellaMorte:2017dyu}.
The RBC/UKQCD group has shown that the GSLM method gives
$30\%$ larger FV effects~\cite{Meyer:2018PoS,Lehner:2018PoS} than the 2-pion-XPT for $(L_1,L_2) = (4.66, 6.22)$ fm.
The ETM collaboration invented the dual QCD parameterization for the light-quark correlator
in the intermediate scale~\cite{Giusti:2018mdh}
and its contribution to the FV effect is investigated in addition to the long distance contributions.
This gives two times larger FV effects than the 2-pion-XPT estimate.
Finally, the direct LQCD estimate gives even larger FV effects than any of aboves:
$\Delta\amuudlohvp(5.4~\text{fm},10.8~\text{fm}) = 40(18)$~\cite{Shintani:2018PoS}
though the statistical error is large and has overlap to the other results.
Thus, the FV effect tends to become larger by taking account of various mode contributions besides
the lowest pion mode.
\footnote{
In the 2-pion-XPT estimate for FV, 100\% error-bar has been assigned, conservatively enough~\protect\cite{Borsanyi:2017zdw}.}

\subsection{Extrapolation to continuum limit and physical mass point}\label{subsec:continuum}
In LQCD, simulations are carried out with finite lattice spacings ($a$) and bare quark masses
which do not reproduce the exact physical meson mass spectra at simulation points.
It is challenging to control a continuum and a physical mass point extrapolations.

\begin{figure}[t]
\vspace{-2mm}
\begin{center}
\includegraphics[width=0.5\textwidth]{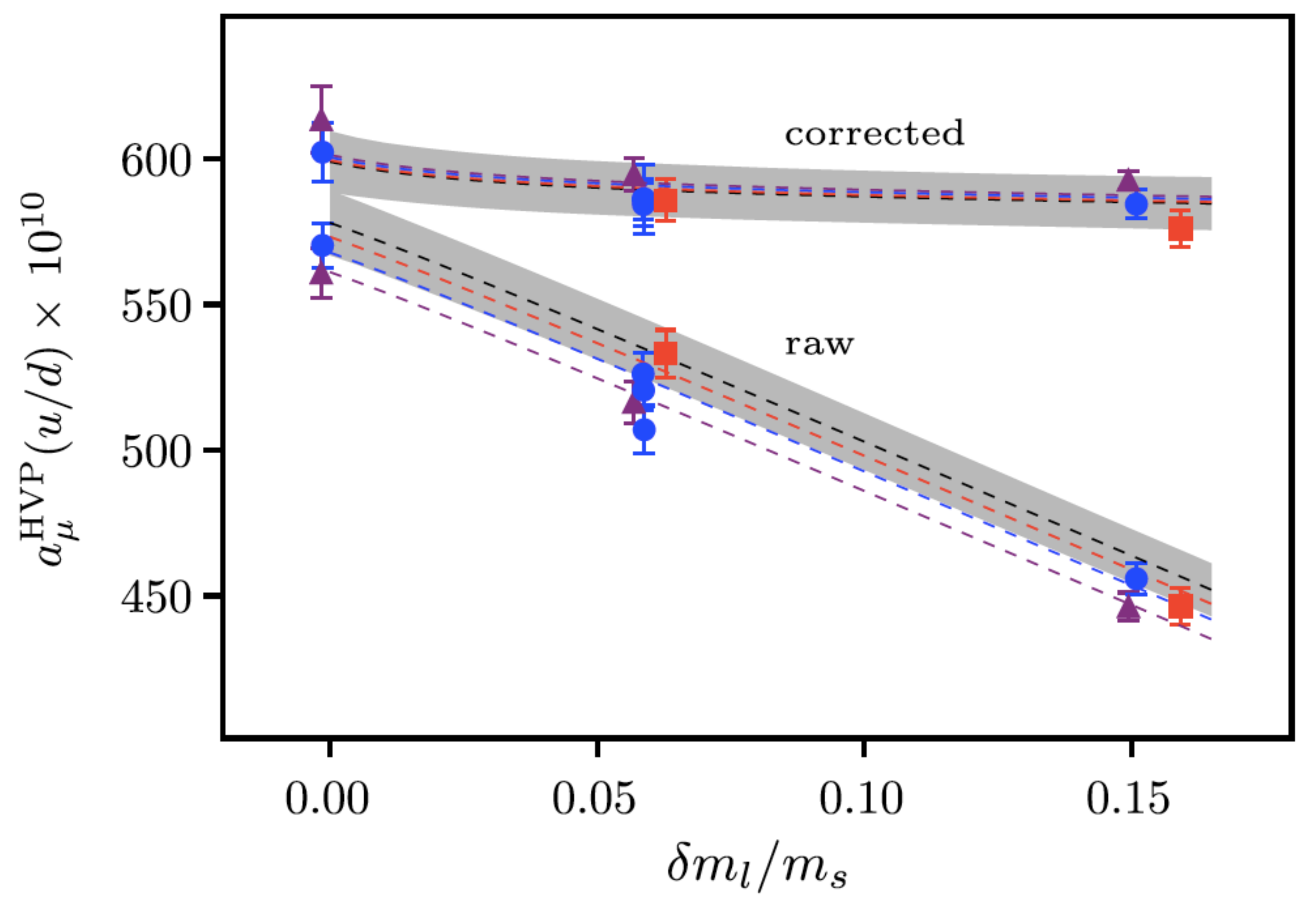}
\includegraphics[width=0.49\textwidth]{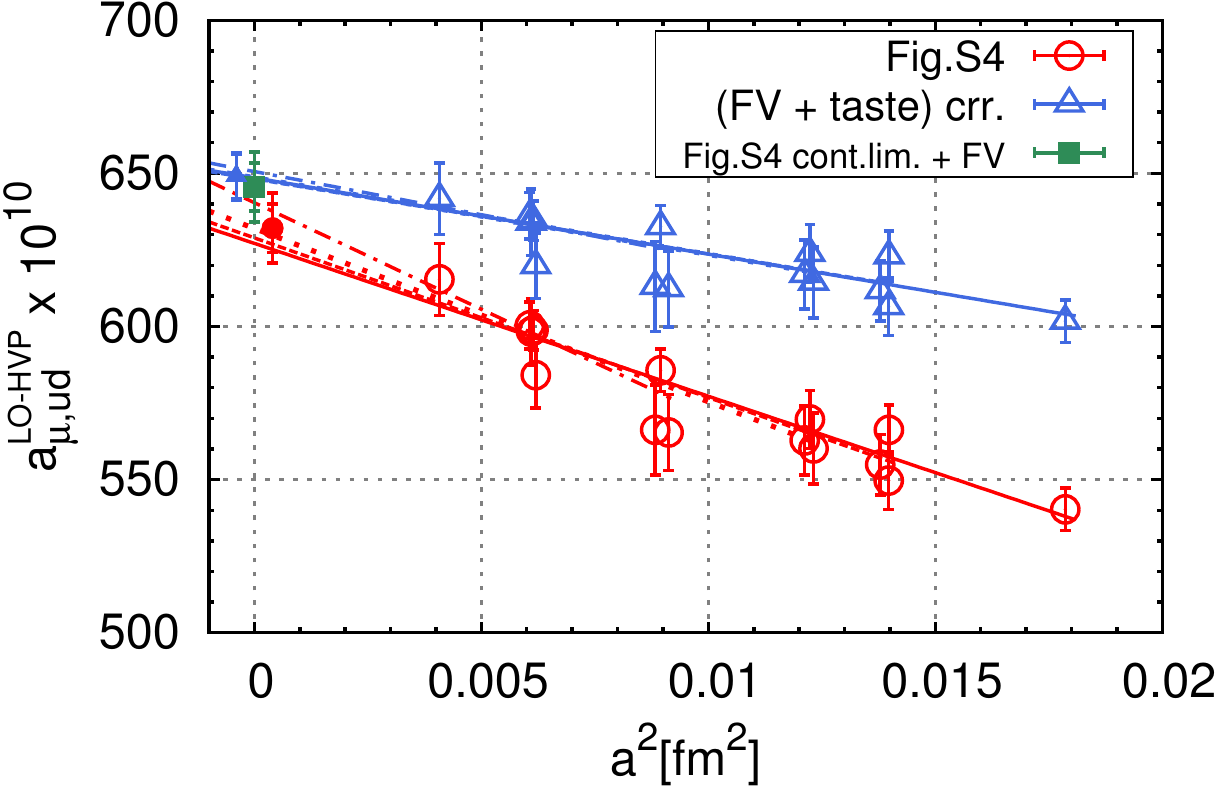}
\caption{
Left: The physical mass point extrapolation by HPQCD collaboration, quoted from Ref.~\protect\cite{Chakraborty:2016mwy}.
Right: The continuum extrapolation by BMW collaboration~\protect\cite{Borsanyi:2017zdw}. See text for details.
}\label{fig:ud_cntxtr}
\end{center}
\vspace{-4mm}
\end{figure}

HPQCD collaboration developed a pion-rho meson effective theory and
derived formulas to calculate the lattice spacing ($a$), taste breaking, and FV effects
in the arbitrary order of the moments (\ref{eq:Pin}).
Using the Pad\'{e} approximants where the moments are the key quantities,
they investigated the extrapolations of $\amulohvp$ to the continuum and physical mass point.
The left panel of Fig.~\ref{fig:ud_cntxtr} shows the physical mass point extrapolation of
the light quark contribution to the anomaly $\amuudlohvp$
by the HPQCD~\cite{Chakraborty:2016mwy}. The upper (lower) curves show their corrected (uncorrected) data:
$N_f = $ 2+1+1 HISQ ensembles,
$a = 0.15$ (purple triangles), 0.12 (blue circles), and 0.09 fm (red squares).
The corrected data become almost flat and stable against the extrapolation.

BMW collaboration performed simulations using $O(10^2 - 10^4)$ stochastic source measurements
with various lattice spacings ranging from $0.064$ fm to $0.134$ fm
in a large volume $m_{\pi}L\gtrsim 4$ at {\em almost physical quark masses} for all ensembles
($N_f = $ 2+1+1 staggard quarks, isospin limit, without QED)~\cite{Borsanyi:2017zdw,Borsanyi:2016lpl}.
With those high quality ensembles,
it becomes possible to perform {\em linear interpolations} to the physical mass point
(rather than the extrapolation) and well-controlled continuum extrapolations
not only for quark-connected contributions (up/down, strange, charm),
but also for up/down/strange-disconnected contributions.
The fit functions are,
\begin{align}
%
&F_{\rm {ud/disc}}
= \amuuddisclohvp(1 + A_2^{\rm ud/disc}a^2)\ ,\quad
F_{\rm s}
= \amuslohvp(1 + A_2^sa^2) + C_K\delta \bar{m}_K^2\ ,\nn\\
&F_{\rm c}
= \amuclohvp(1 + A_2^ca^2) + C_{\pi}\delta m_{\pi}^2 + C_{\eta_c} \delta m_{\eta_c}\ ,\label{eq:fit}
\end{align}
where
$\delta m_{\pi}^2 = m_{\pi}^2 - (m_{\pi}^{\text{flag-isl}})^2$,
$\delta \bar{m}_{K}^2 = \bar{m}_{K}^{2} - (\bar{m}_{K}^{\text{flag-isl}})^2$
with $\bar{m}_K^2 = m_{K}^2 - m_{\pi}^2/2$, and
$\delta m_{\eta_c} = m_{\eta_c} - m_{\eta_c}^{\text{hpqcd}}$.
The suffix ``flag-isl'' indicates the isospin-limit value reported by the FLAG collaboration~\cite{Aoki:2016frl},
and ``hpqcd'' is the value reported by HPQCD collaboration~(Appendix of Ref.~\cite{Chakraborty:2014aca})
in which the QED subtraction is taken care.
The fit parameter $a_{\mu,\cdots}^{\text{{\tiny LO-HVP}}}$ in Eq.~(\ref{eq:fit}) is interpreted
as each flavor contributions to the total $\amulohvp$
at the continuum limit and the physical point up to SIB/QED corrections, which will be separately evaluated in later.
The $\chi^2$, for example in the case of strange contributions, is given by
$\chi^2 = {}^t\Delta C^{-1} \Delta$ where
${}^t\Delta = (a - a^{\rm dat},\bar{m}_K - \bar{m}_K^{\rm dat}, F_s - a_{\mu,s,{\rm dat}}^{\text{{\tiny LO-HVP}}})$
and $C$ represents the covariance matrix constructed by
$(a^{\rm dat},\bar{m}_K^{\rm dat},a_{\mu,\text{f = s,dat}}^{\text{{\tiny LO-HVP}}})$.
We note that ($a,\bar{m}_K$) in the $\chi^2$ and the fit function $F_s$ denote fit parameters and
not the lattice data themselves, and thus full correlations are taken account.
The $\chi^2$ for the other flavors are constructed similarly,
and a good fit quality is achieved for all connected and disconnected contributions.
\footnote{The fit functions may be modified from Eq.~(\protect\ref{eq:fit})
in different LQCD groups;
\textcircled{\scriptsize 1} if non-staggard fermions without $\mathcal{O}(a)$ improvement are used,
the fit function should include a term linearly dependent on $a$,
\textcircled{\scriptsize 2} higher order terms of $a$ might be necessary,
\textcircled{\scriptsize 3} pion mass correction terms, (e.g. $\delta m^2_{\pi}$ and/or
$m_{\pi}^2\log m_{\pi}^2 - (m_{\pi}^{\text{flag-isl}})^2\log (m_{\pi}^{\text{flag-isl}})^2$),
might be necessary,
\textcircled{\scriptsize 4} if ensembles do not include charm sea-quarks,
the $\Delta m_{\eta_c}$ term in $F_c$ becomes irrelevant.}

In the right panel in Fig.~\ref{fig:ud_cntxtr},
we show the continuum extrapolations by BMW collaboration~\cite{Borsanyi:2017zdw}.
The red-open-circles represent $\amuudlohvp$ without FV/taste corrections
and are continuum-extrapolated to the red-filled-circle, for which FV effects estimated by 2-pion-XPT are added
to get the final estimate (green-square, Fig.~S4 in the supplemental material of Ref.~\cite{Borsanyi:2017zdw}).
This result is well agree to the blue-filled-triangle,
which is obtained through the different procedure from the green-square;
similarly to the HPQCD method~\cite{Chakraborty:2016mwy},
the FV and taste corrections estimated by the XPT with pions and rho-mesons
are added to the red-open-circles and then continuum-extrapolated to get the blue-filled-triangle.
The good agreement indicates the reliability of the results.
The statistical and systematic uncertainties in the light component continuum extrapolations are both
$\sim 1.2$\%, which must be further reduced to achieve the target precision.

\subsection{Strong-Isospin Breaking and QED corrections}\label{subsec:SIB_QED}
For the target precision $\mathcal{O}(0.1\%)$,
the strong-isospin breaking (SIB, $\mathcal{O}(\delta m /\Lambda_{\rm QCD})\sim 1\%$,
$\delta m = m_d - m_u = 2.41(6)(4)(9)$ MeV in $\overline{\text{MS}}$[2 GeV]~\cite{Fodor:2016bgu})
and QED ($\mathcal{O}(\alpha)\sim 1\%$) corrections must be taken account.
For SIB, three different strategies have been considered;
\textcircled{\scriptsize 1}
direct LQCD simulations where
the input up and down quark masses set different,
\textcircled{\scriptsize 2}
use isospin symmetric ensembles and the SIB effects are encoded in the valence quarks
in observable measurements,
\textcircled{\scriptsize 3}
use isospin symmetric ensembles and perform
a perturbative expansion in terms of $\delta m = m_d - m_u$~\cite{deDivitiis:2011eh}.
\begin{wrapfigure}{r}{0.45\textwidth}
\vspace{-4mm}
\hspace{-4mm}
\begin{center}
\includegraphics[width=0.45\textwidth]{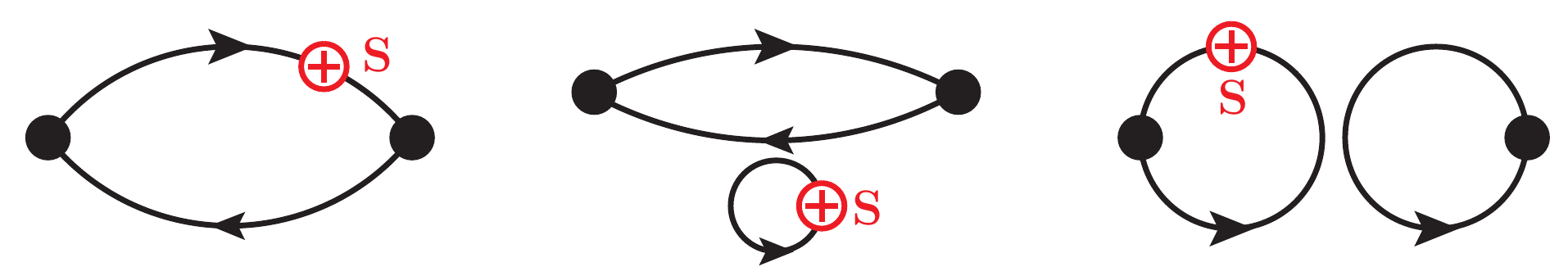}
\includegraphics[width=0.45\textwidth]{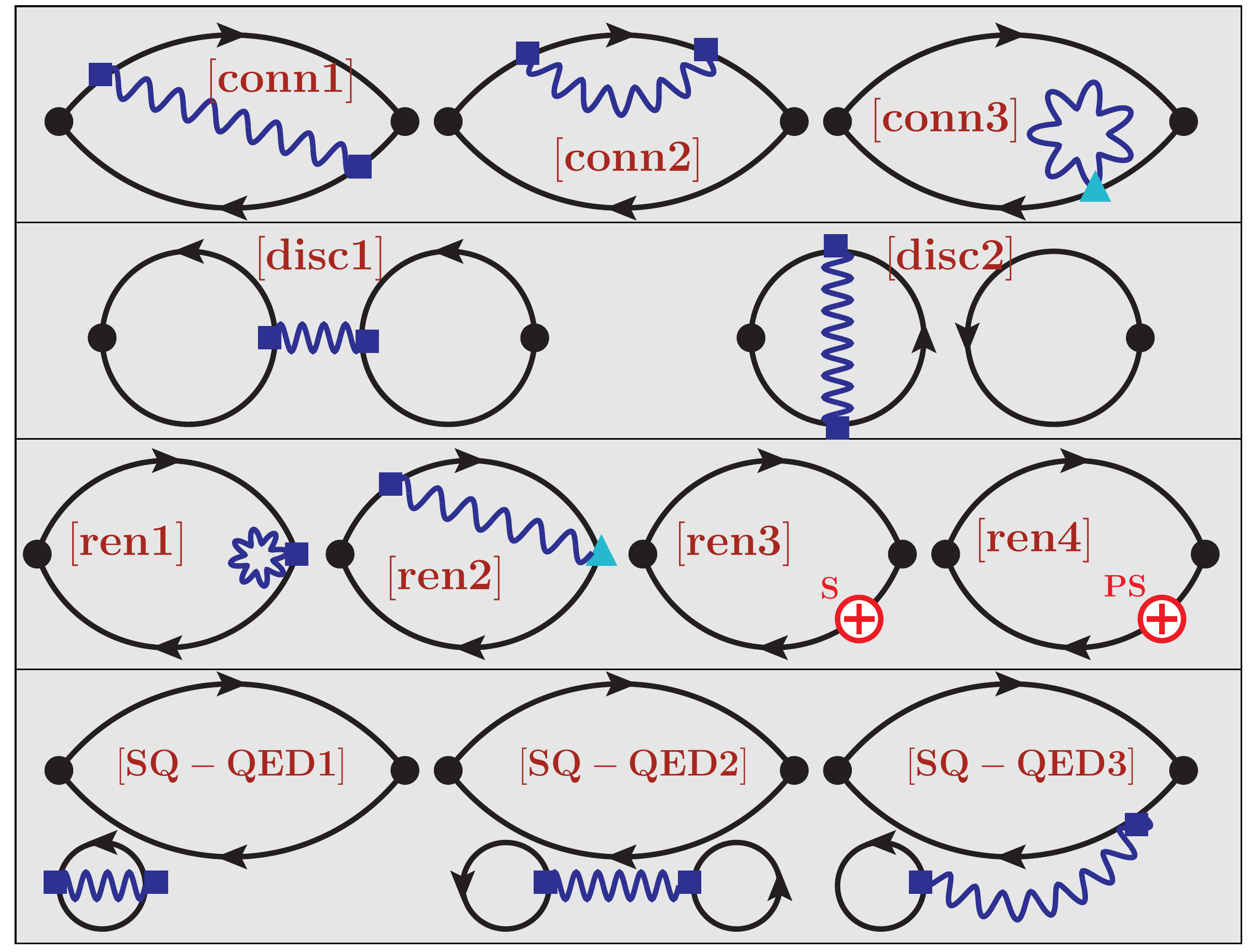}
\caption{
Upper: SIB corrections.
{\color{red} $\bigoplus$} = scalar ($S_{\rm br}$) insertion
in the quark propagator.
Lower: QED corrections in the perturbative method.
The wavy lines represent the inserted photons.
{\color{indigo-dye} $\blacksquare$} =  vector-current,
{\color{darkspringgreen} $\blacktriangle$} =  tadpole,
{\color{red} $\bigoplus$} =  (pseudo-)scalar insertions.
From the first line, quark-connected, quark-disconnected, renormalization, and sea-quark QED corrections.
}\label{fig:sib}
\end{center}
\vspace{-4mm}
\end{wrapfigure}
The first and second methods were recently adopted by FHM collaboration~\cite{Chakraborty:2017tqp};
the muon anomaly $\amulohvp$ was calculated for both $N_f$ = (1+1+1+1) and (2+1+1) ensembles
with almost physical up and down quark masses,
and a valence quark mass ($m_{\rm val}$) dependence of $\amulohvp$ was compared between them.
Two results agree at {$m_{\rm val} = m_l = (m_u + m_d)/2$},
which indicates the SIB effects in sea-quarks (included in \textcircled{\scriptsize 1} only) are negligible.
The SIB correction is then approximated by
{$\delta a_{\mu}^{\text{{\tiny LO-HVP}}} =
(q_u^2 a_{\mu}^{\text{{\tiny LO-HVP}}}|_{m_u}
+ q_d^2a_{\mu}^{\text{{\tiny LO-HVP}}}|_{m_d})
- (q_u^2 + q_d^2)a_{\mu}^{\text{{\tiny LO-HVP}}}|_{m_l}$}
where the charge factors  $q_{u,d}$ are defined in Eq.~(\ref{eq:jmu}).
The result by FHM group~\cite{Chakraborty:2017tqp} is:
{$\delta a_{\mu}^{\text{{\tiny LO-HVP}}} / a_{\mu}^{\text{{\tiny LO-HVP}}} = 1.5(7)\% $} (positive).

We shall move on the perturbative method \textcircled{\scriptsize 3}~\cite{deDivitiis:2011eh}.
Let $S$ be the QCD action with SIB effects.
The up/down mass terms $(m_u\bar{u}u + m_d\bar{d}d)$ in $S$ is rewritten as
$m_l(\bar{u}u + \bar{d}d) - \delta m(\bar{u}u - \bar{d}d)$ with $m_l = (m_u + m_d)/2$ and $\delta m = (m_d - m_u)/2$.
Therefore, one find $S = S_0 - \delta m S_{\rm br}$ where $S_0$ denotes the isospin symmetric part
and $S_{\rm br} = \sum_x(\bar{u}u - \bar{d}d)(x)$ represents the breaking term.
Consider the ensemble average of an operator $O$ and expand it in terms of $\delta m$:
$
\langle O \rangle
= \int O e^{-S} / \int e^{-S}
\simeq \int O (1 + \delta m S_{\rm br})e^{-S_0} / \int (1 + \delta m S_{\rm br})e^{-S_0}
\simeq \langle O \rangle_0 + \delta m \langle OS_{\rm br}\rangle_0\ ,
$
where $\langle\cdot\rangle_0$ represents
the ensemble average with the isospin symmetric weight $e^{-S_0}$.
This method allows us to reuse existing isospin-symmetric ensembles to evaluate the SIB effect.
In the present context, $O = G_{\mu\nu}(t)$ (c.f. Eq.~(\ref{eq:Pimn})),
and the Wick theorem tells us that
the SIB term $\delta m\langle G_{\mu\nu}(x)S_{\rm br}\rangle_0$ produces
the diagrams shown in the upper panel of Fig.~\ref{fig:sib}.
RBC/UKQCD group has shown that the perturbative method
is consistent~\cite{Boyle:2017gzv} with the valence quark method \textcircled{\scriptsize 2}
and results in the 1.5(1.2)\% positive correction to the total $\amulohvp$
at the physical mass point~\cite{Blum:2018mom}, consistently to the FHM results mentioned above.


We shall now investigate the (QCD + QED) system for which the partition function reads,
$
\langle O \rangle
= (1/Z)\int_{q,\bar{q},U}\int_A O
e^{-S_F[q,\bar{q},U,A] - S_G[U] - S_{\gamma}[A]},
$
with $O = G_{\mu\nu}(t)$ (c.f. Eq.~(\ref{eq:Pimn})) in the present context.
To ensure the transfer matrix well-defined,
the non-compact QED ($S_{\gamma}[A] = (a^4/4)(\sum_x\partial_{\{\mu}A_{\nu\}})^2$)
is usually adopted in the Coulomb gauge ($\vec{\partial}^{\dagger}\cdot\vec{A} = 0$).
To control QED FV effects, $\text{QED}_L$ prescription \cite{Hayakawa:2008an} is used;
spatial zero-modes and the universal {$1/L^{n = 1,2}$} corrections to mass are removed,
while a reflection positivity is preserved.

There are several strategies to include the QED effects;
\textcircled{\scriptsize 1}
full (QCD + QED) simulations,
which is, at this moment, available for $m_{\pi} = 400$ MeV case only~\cite{Zanotti:2018PoS},
\textcircled{\scriptsize 2}
stochastic method~\cite{Duncan:1996xy} where the photon fields $A_{\mu}$ in the quark determinant are ignored
and stochastically generated with weight $e^{-S_{\gamma}}$
independently of gluon fields $U_{\mu}$ (electro-quenched), and multiplied,
$U_{\mu}(x)\to e^{-ieq_f A_{\mu}(x)}U_{\mu}(x)$,
\textcircled{\scriptsize 3}
perturbative method where QED is treated in the perturbative expansion
in $\alpha = e_{\rm el}^2/(4\pi)$~\cite{deDivitiis:2013xla},
$\langle G_{\mu\nu} \rangle
= \langle G_{\mu\nu} \rangle_0
+ \frac{e^2_{\rm el}}{2}\frac{\partial^2 \langle G_{\mu\nu} \rangle}{\partial e^2_{\rm el}}\Big|_{e_{\rm el} = 0}
+ \mathcal{O}(\alpha^2)$,
where $\langle\cdot\rangle_0 = \langle\cdot\rangle|_{e_{\rm el} \to 0}$.
The derivative $\partial \langle\cdot \rangle/ \partial e_{\rm el}$
picks up a combination of a photon $A_{\mu}$ and a vector current operators $j_{\mu}$
from $e^{-S_F[q,\bar{q},U,A]}$ and $e_{\rm el}\to 0$ taken after.
Therefore, the QED corrections are expressed as $A_{\mu}$ and $j_{\mu}$ insertions
to the quenched average $\langle G_{\mu\nu}\rangle_0$.
In additions, a mass retuning due to QED inclusion is necessary via the scalar operator insertion.
If the twisted mass boundary condition is used, the pseudo-scalar operator insertion is also necessary
to keep the twist originally set.
Diagrams with the various insertions are displayed in the lower panel of Fig.~\ref{fig:sib}.
\footnote{Besides the diagrams shown in Fig.~\protect\ref{fig:sib},
the QED corrections to the strong coupling constant would be necessary.
See Ref.~\protect\cite{Risch:2018ozp} for details.}
The stochastic and perturbative methods gave consistent corrections~\cite{Boyle:2017gzv}.

\begin{wrapfigure}{r}{0.57\textwidth}
\vspace{-6mm}
\hspace{-4mm}
\begin{center}
\includegraphics[width=0.57\textwidth]{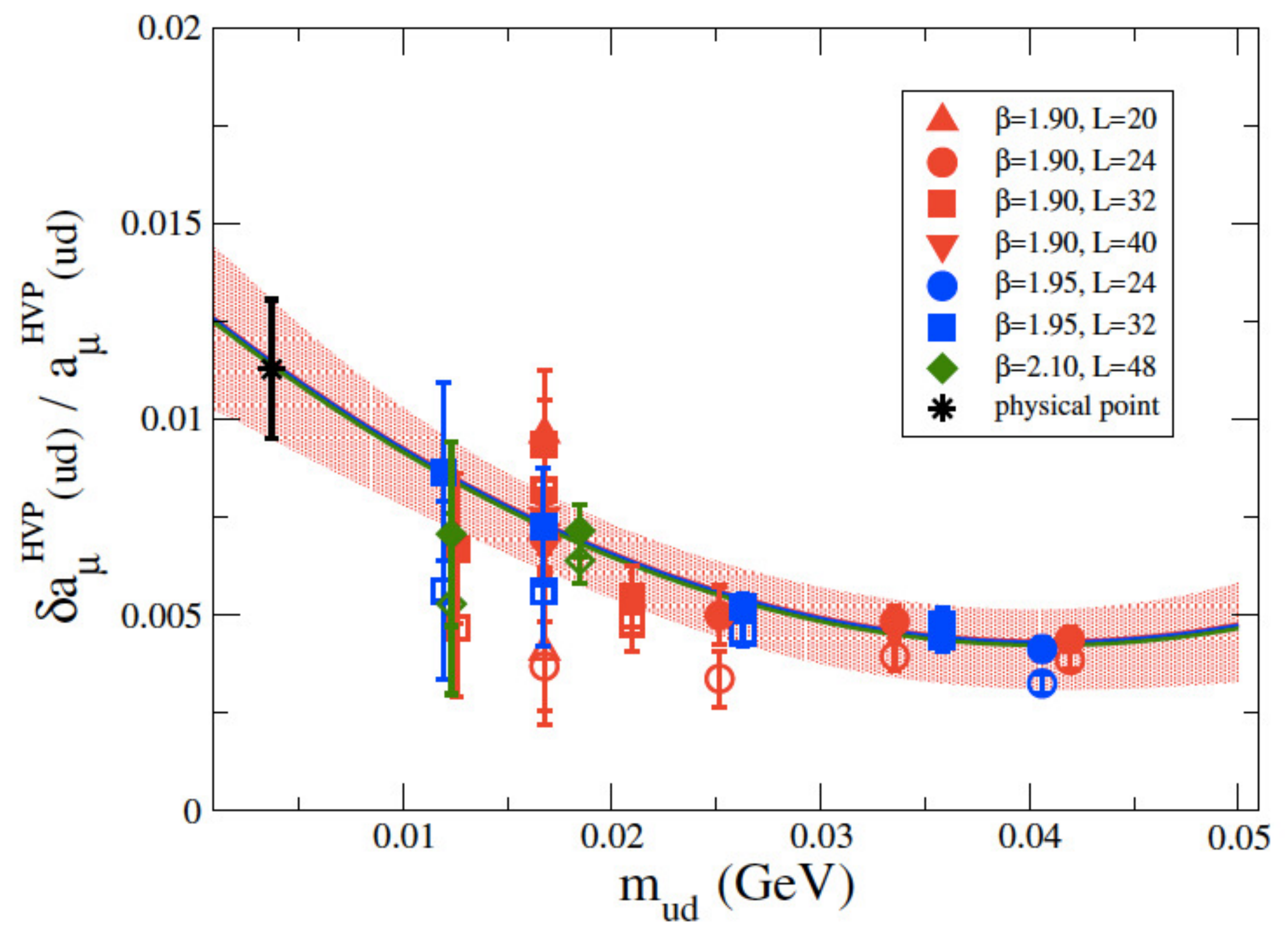}
\caption{
Quoted from Ref.~\protect\cite{Giusti:2018vrc}:
SIB and QED corrections in the light quark anomaly by ETM group.
See text for details.
}\label{fig:qed}
\end{center}
\vspace{-4mm}
\end{wrapfigure}
The most precise estimates for the SIB and QED corrections have been reported
from the ETM group~\cite{Giusti:2018vrc} where the perturbative method
explained above~\cite{deDivitiis:2011eh,deDivitiis:2013xla} have been adopted.
Figure~\ref{fig:qed} shows the ETM results
on the corrections to the light quark anomaly $\amuudlohvp$ as a function of the renormalized mass.
The open and filled symbols corresponds to the raw and FV-corrected LQCD data, respectively.
The solid line shows the fit for the corrected data and gives the correction at the physical point (black asterisk),
which is around 1\% correction with a few per-mil uncertainty ($\delta \amulohvp\times 10^{10} = 7(2)$).
In Table~\ref{tab:sib_qed}, the SIB/QED corrections obtained via various methods are compared.
All results are $\sim \mathcal{O}(1) \%$ and consistent within relatively large error-bars,
which must be reduced for the target precision $\mathcal{O}(0.1)$ \% in future.

\subsection{Other developments}\label{subsec:others}
There are many subjects which are related to the HVP and $\amulohvp$ but omitted in this proceedings.
See the following references:
LQCD study on the isospin breaking in tau decay and muon $g-2$~\cite{Bruno:2018ono},
LQCD combined with MUonE experiments~\cite{Marinkovic:2018PoS,Giusti:2018mdh},
NLO-HVP contributions to muon $g-2$~\cite{Chakraborty:2018iyb},
LO-HVP contribution to the weak-mixing angle~\cite{Ce:2018ziv} and the dark photon search~\cite{Pospelov:2008zw},
and CKM matrix element $|V_{us}|$ from LQCD (V + A) current correlators~\cite{Boyle:2018ilm}.

\section{Comparison and Discussion}\label{sec:discuss}
This section is devoted to show $\amulohvp$ reported by various LQCD groups and compare them.
The combined results using the LQCD and the dispersive method are also discussed.
The $\amulohvp$ to be compared takes account of the extrapolations to the continuum limit and the physical mass point,
and FV/SIB/QED corrections.
The uncertainties include a statistical error
and systematic errors from
a scale setting, lattice data cuttings, fit model dependences in the extrapolations/interpolations,
IR-cuts in the correlators $(C^{f=ud},D)(t)$, and/or UV-cuts in the HVP $\hat{\Pi}(\omega^2)$.
Both statistical and total systematic errors are at a few percent level at present.

\subsection{Comparing LQCD results}\label{subsec:compare_lat}
%
\begin{wrapfigure}{r}{0.5\textwidth}
\vspace{-4mm}
\hspace{-4mm}
\begin{center}
\includegraphics[width=0.5\textwidth]{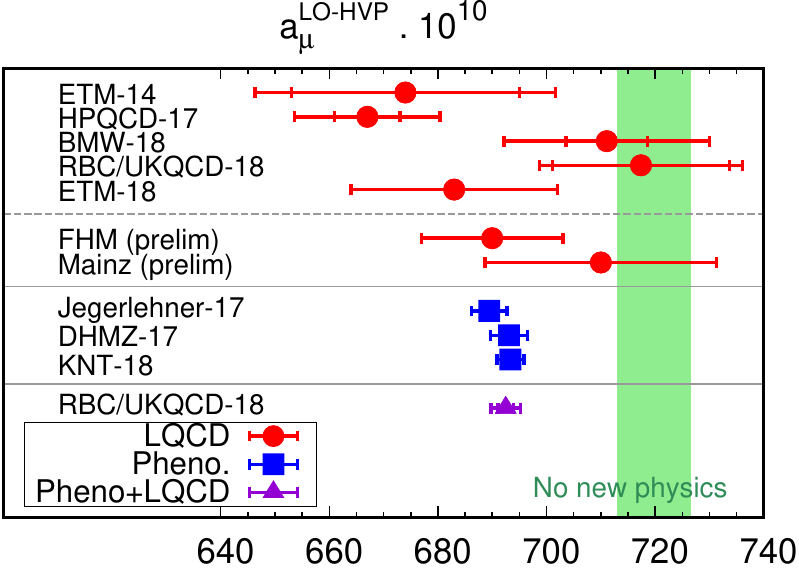}
\caption{
Compilation of recent results for the $\amulohvp$ in units of $10^{-10}$.
See text for details and Table~\protect\ref{tab:amu_cmp}.
}\label{fig:amu_compare}
\end{center}
\vspace{-4mm}
\end{wrapfigure}
In Fig.~\ref{fig:amu_compare} and Table~\ref{tab:amu_cmp}, we compare $\amulohvp$ reported by various LQCD groups
as well as the one from the dispersive method.
The recently published results, BMW-18~\cite{Borsanyi:2017zdw}
and RBC/UKQCD-18~\cite{Blum:2018mom}, are consistent well to each other and
no new physics (green band in the figure): the value that $\amulohvp$ would have to explain
the experimental measurement of $a_{\mu}$~\cite{Bennett:2006fi}, assuming that all other SM contributions are unchanged.
In contrast, HPQCD-17~\cite{Chakraborty:2016mwy}, ETM-14~\cite{Burger:2013jya}, and ETM-18~\cite{Giusti:2018mdh}
have observed a smaller $\amulohvp$ than no new physics.
Recently, HPQCD-17 is updated to FHM-prelim., which becomes closer to the BMW-18 and RBC/UKQCD-18 estimates.
All (updated) results are consistent with the dispersive estimates
where the latter uses Eq.~(\ref{eq:disp}) to calculate the HVP.
Thus, the present LQCD estimates of $\amulohvp$ are still premature to confirm or infirm
the deviations among the experimental measurement and the dispersive SM predictions.

\begin{figure}[!t]
\vspace{-2mm}
\begin{center}
\includegraphics[width=0.48\textwidth]{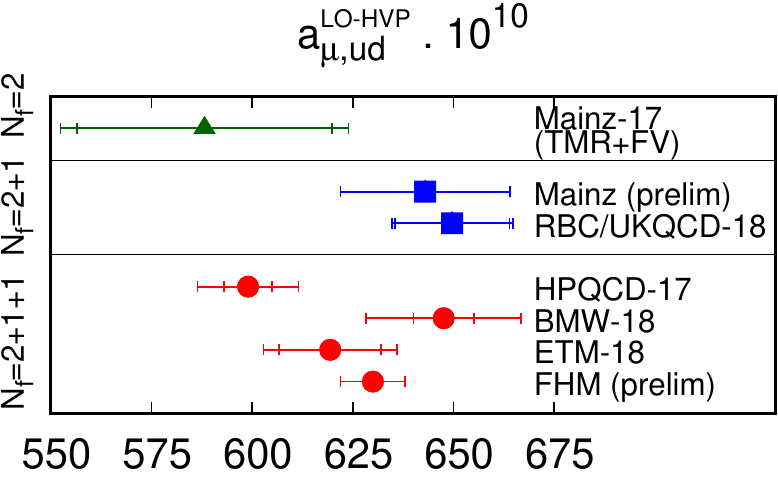}
\includegraphics[width=0.48\textwidth]{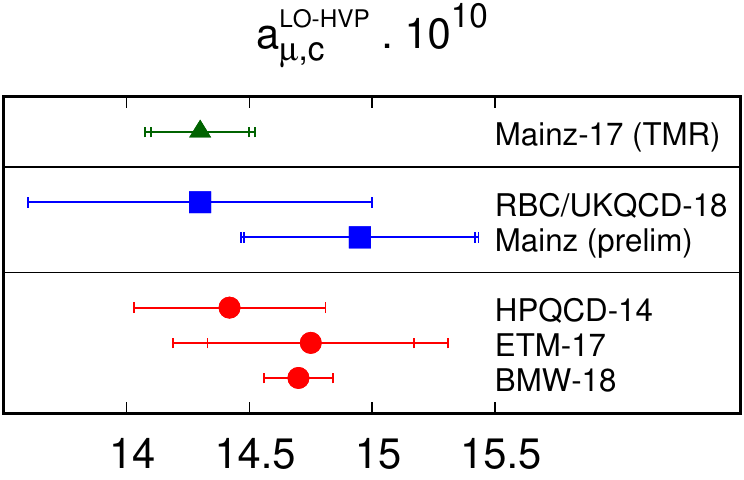}

\includegraphics[width=0.48\textwidth]{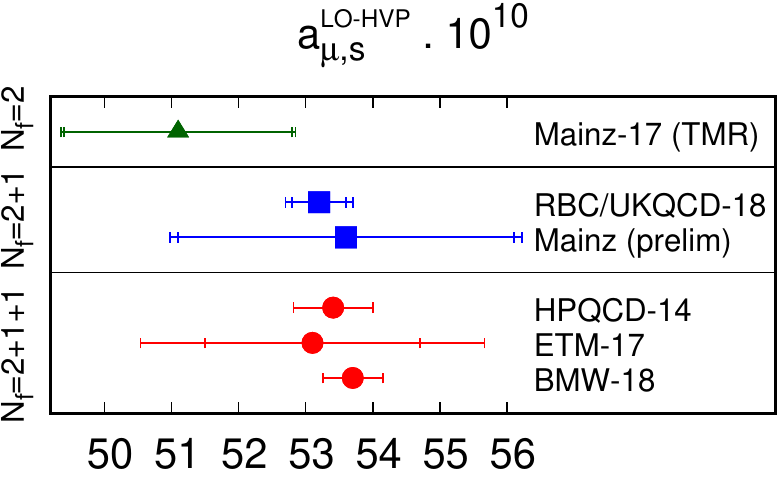}
\includegraphics[width=0.48\textwidth]{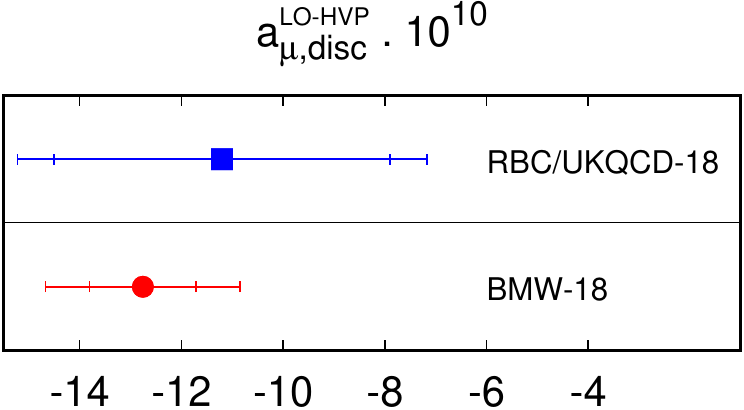}
\caption{
Comparison in flavor-by-flavor:
connected light/strange/charm contributions ($a_{\mu,ud/s/c}^{\text{{\tiny LO-HVP}}}$, upper-left/lower-left/upper-middle)
and disconnected contributions ($\amudisclohvp$, lower-middle).
}\label{fig:amu_compare_fbyf}
\end{center}
\vspace{-4mm}
\end{figure}

As seen in Fig.~\ref{fig:amu_compare}, the LQCD published results are not fully consistent to each other.
To see how the tension comes out, we compare $\amulohvp$
in flavor-by-flavor in Fig.~\ref{fig:amu_compare_fbyf} (see also Table.~\ref{tab:amu_fbf}):
connected light/strange/charm contributions ($a_{\mu,ud/s/c}^{\text{{\tiny LO-HVP}}}$, upper-left/lower-left/upper-right)
and disconnected contributions ($\amudisclohvp$, lower-right).
The $a_{\mu,s/c/\text{disc}}^{\text{{\tiny LO-HVP}}}$ are already determined
with high enough precision with respect to the requirements from FNAL-E989 and J-PARC-E34 experiments
and consistent among all LQCD groups. The tension is on the light connected contribution $\amuudlohvp$
in the published results as shown in the upper-left panel.
\footnote{
The first and second moments defined in Eq.~(\protect\ref{eq:Pin}) are also indicative of how the tension comes out.
See Table.~\protect\ref{tab:tm_fbf}.}
\footnote{
It should be noted that the discrepancy in the $\amuudlohvp$ between HPQCD-17 and the others
(upper-left panel in Fig.~\ref{fig:amu_compare_fbyf}) is somewhat overestimated;
In HPQCD-17,
FV and taste-breaking corrections are calculated in the framework of the $\pi - \rho$ effective model where
the corrections associated with the {\em disconnected} $\pi-\pi$ diagrams has not been excluded.
If this correction is excluded, the results would become somewhat larger.}
FHM collaboration has updated their ensembles and improved the multi-exponential fits
for the light quark connected correlator $C^{ud}(t)$ at large distance,
which modified their result to $\amuudlohvp = 630(8)$~\cite{FHM:private-discuss}.
In turn, $\amuudlohvp$ by RBC/UKQCD tends to become smaller
when the higher excitation modes are taken account in the improved bounding method
for $C^{ud}(t)$ at large distance~\cite{Meyer:2018PoS}.
Thus, the tension in $\amuudlohvp$ seems related to the treatment of the long distance behavior in $C^{ud}(t)$
and relaxed in the updated results.

\subsection{Lattice combined with dispersive method}\label{subsec:lat_pheno}

\begin{figure}[!t]
\vspace{-2mm}
\begin{center}
\includegraphics[width=0.495\textwidth]{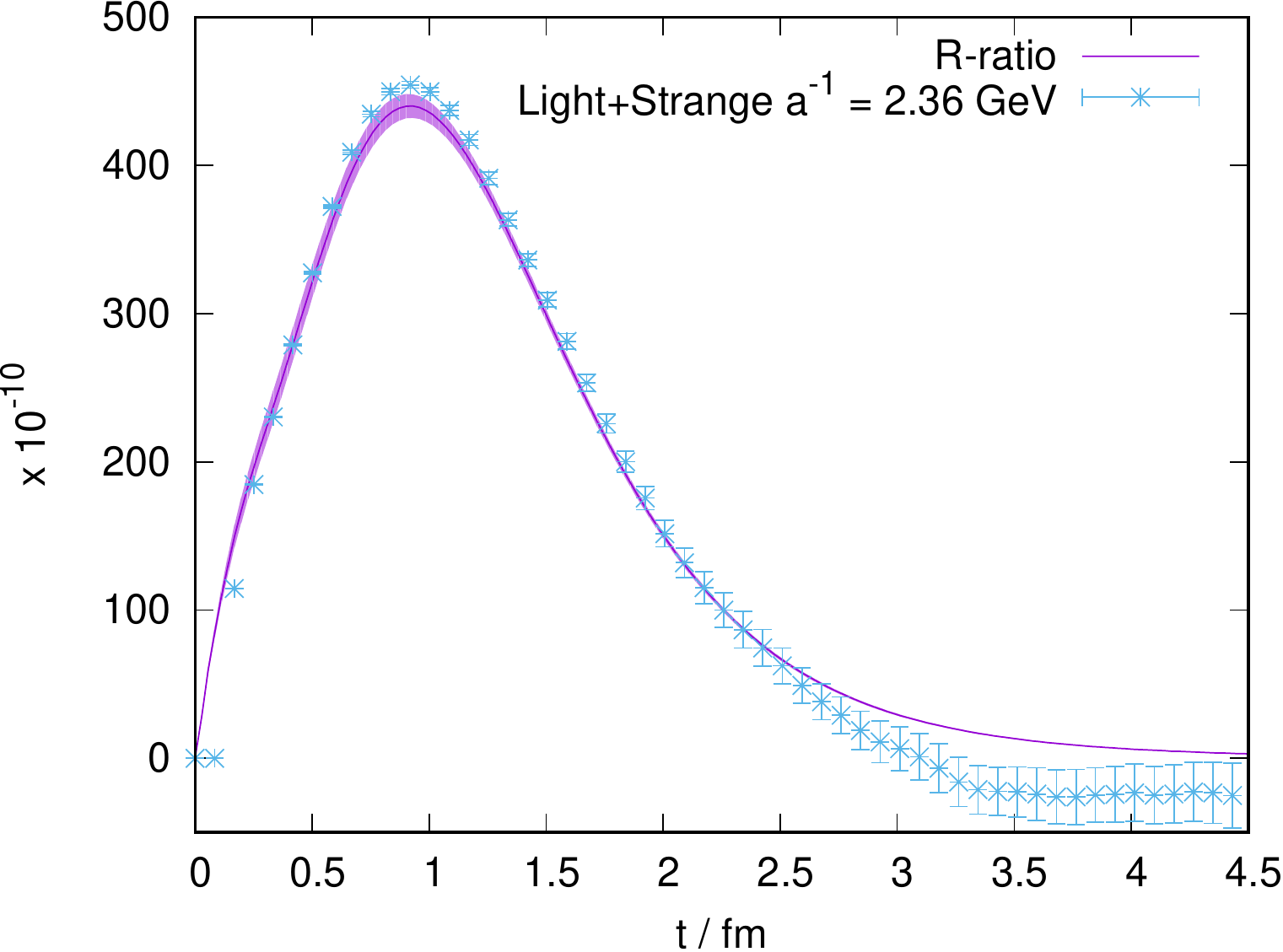}
\includegraphics[width=0.495\textwidth]{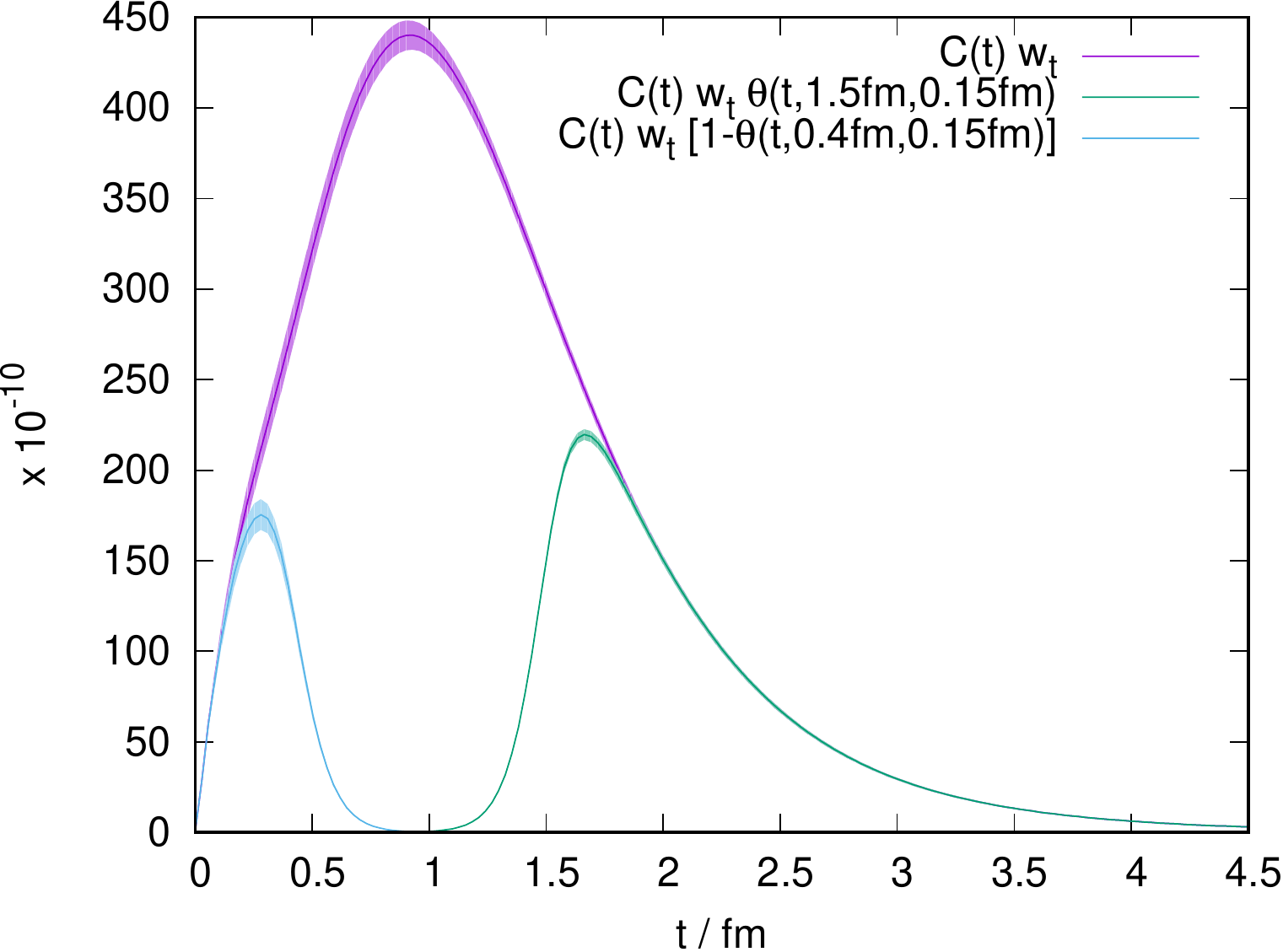}
\caption{
The window method. Figures are quoted from Ref.~\protect\cite{Blum:2018mom}.
Left:
The vector current correlator multiplied by weight factor ($(\alpha/\pi)^2 W(t,Q^2/m_{\mu}^2) G(t)$)
is compared between the LQCD and dispersive method.
Right:
The weighted correlator by dispersive method $(\alpha/\pi)^2 W(t,Q^2/m_{\mu}^2) G_{\text{ph}}(t)$ (purple)
and the partial contributions from short-distance (SD, light-blue) and long-distance (LD, green).
For intermediate distance, the LQCD result will be adopted instead of the dispersive one.
}\label{fig:window_rbcuk}
\end{center}
\vspace{-4mm}
\end{figure}

The combination of the LQCD and dispersive method may deliver the estimates of $\amulohvp$
at a per-mil level close upon the target precision.
The left panel of Fig.~\ref{fig:window_rbcuk} shows the integrand in Eq.~(\ref{eq:tmr_a})
for LQCD data by RBC/UKQCD~\cite{Blum:2018mom} and the dispersive method~\cite{jeg:alphaQEDc17}:
$G_{\rm lat/ph}(t)W(t,m_{\mu})$.
The suffix ``lat'' and ``ph'' show LQCD and phenomenological vector current correlators, respectively,
where the latter is defined via the R-ratio (c.f. Eq.~(\ref{eq:disp})),
$G_{\rm ph}(t) = (1/2)\int_0^\infty ds \sqrt{s}\frac{R(s)}{3}e^{-\sqrt{s}|t|}$.
In the left panel, a peak is around $t\sim 1$ fm,
where the LQCD data would not suffer from the discretization nor FV artifacts and are determined more precisely
than the dispersive method. RBC/UKQCD group has proposed~\cite{Blum:2018mom} that
the $G_{\rm lat}(t)$ is used for that region ($t\sim 1$ fm) while the dispersive data
are adopted for the UV and IR regions. Consider the decomposition,
\begin{align}
& G_{\rm lat/ph}(t) = (G_{\rm lat/ph}^{\text{SD}} + G_{\rm lat/ph}^{\text{W}} + G_{\rm lat/ph}^{\text{LD}})(t)\ ,\\
& (G_{\rm lat/ph}^{\text{SD}},G_{\rm lat/ph}^{\text{W}},G_{\rm lat/ph}^{\text{LD}})(t) =
G_{\rm lat/ph}(t)(1 - \Theta(t,t_0,\Delta),\Theta(t,t_0,\Delta) - \Theta(t,t_1,\Delta),\Theta(t,t_1,\Delta))\ ,
\end{align}
with the smeared step function, $\Theta(t,t^{\prime},\Delta) = 0.5(1 + \tanh[(t-t^{\prime})/\Delta])$.
%
%
The light-blue and green lines in the right panel of Fig.~\ref{fig:window_rbcuk}
represent $G_{\rm ph}^{\text{SD}}(t,\Delta)W(t,m_{\mu})$ and
$G_{\rm ph}^{\text{LD}}(t,\Delta)W(t,m_{\mu})$, respectively.
The missing contribution (difference between (light-blue + green) and purple lines) is compensated
by LQCD data $G_{\rm lat}^{\text{W}}(t,\Delta)W(t,m_{\mu})$.
Thus, lat/pheno-combined estimate is
\begin{align}
\frac{a_{\mu,{\rm cmb}}^{\text{{\tiny LO-HVP}}}(t_0,t_1,\Delta)}{(\alpha/\pi)^2} = 
\int_0^{t_0}dt\ {G}_{\rm ph}^{\text{SD}}(t,\Delta)W(t,m_{\mu})
+ \sum_{t = t_0}^{t^1}{G}_{\rm lat}^{\text{W}}(t,\Delta)W(t,m_{\mu})
+ \int_{t_1}^{\infty}dt\ {G}_{\rm ph}^{\text{LD}}(t,\Delta)W(t,m_{\mu})
\ .\nn\label{eq:amu_w}
\end{align}
The window threshold parameters $(t_0,t_1)$ are adjusted to minimize a total uncertainty
and their variation gives a systematic error. The smearing parameter $\Delta$ is adjusted
to control the lattice discretization artifact at the window boundary.
RBC/UKQCD used $(t_0,t_1,\Delta) = (0.4, 1.0, 0.15)$ fm and obtained
$a_{\mu,{\rm cmb}}^{\text{{\tiny LO-HVP}}} = 692.5(2.7)\cdot 10^{-10}$~\cite{Blum:2018mom},
which may be the most precise estimate today.

In Ref.~\cite{Aubin:2018fog}, the window method is considered with the same $(t_0,t_1,\Delta)$ as the above
and the contribution from the intermediate window is evaluated by using two different LQCD ensembles:
HISQ and domain-wall fermions (DWF). In the continuum limit, the DWF becomes consistent with the dispersive method
while the HISQ gets larger than them. The discrepancy is about 2-$\sigma$.
We need more studies on the discretization effects in the window method.

\section{Concluding remarks}\label{sec:conclude}
We have reviewed the LQCD results for HVP and
its leading-order ($\sim\mathcal{O}(\alpha^2)$) contribution to
the muon anomalous magnetic moments $\amulohvp$.
Remarkably enough, more than 3-sigma deviations
between the SM prediction with the QCD dispersion relation used for HVP
and the BNL experiment in 0.5 ppm precision is reported.
This may be a milestone to the BSM physics while it is mandatory to confirm or infirm
the discrepancy based on the ab-initio calculation by the lattice QCD (LQCD) simulations.
In the coming years, FNAL-E989, J-PARC-E34, and MUonE experiments will provide more precise data,
which requires the LQCD to compute $\amulohvp$ with a precision at $\mathcal{O}(0.1)$ \% level.

Significant progress have been made in the LQCD approaches to $\amulohvp$;
1) the noise control in the light connected and quark-disconnected vector current correlators
using stochastic sources with AMA and/or hierarchical technique ({\em c.f.} Sec.~\ref{subsec:target}),
2) the IR behavior of the HVP with model independent approximants ({\em c.f.} Sec.~\ref{subsec:approximant}),
3) the long distance control of the light connected correlator and FV estimates
with GSLM method and many other ideas ({\em c.f.} Sec.~\ref{subsec:long_distance}),
4) the controlled extrapolations to the continuum limit and the physical mass point ({\em c.f.} Sec.~\ref{subsec:continuum}),
5) the various estimates for the SIB/QED corrections ({\em c.f.} Sec.~\ref{subsec:SIB_QED}),
and many others ({\em c.f.} Sec.~\ref{subsec:others}).

The appendix provides the {\em Summary Table 2018} for $\amulohvp$ and related quantities.
Both statistical and total systematic errors in LQCD estimates for $\amulohvp$ are at a few percent level,
which is still much larger than the dispersive estimates.
The large portion of the uncertainty and some disagreements among different groups
are on the light connected contributions $\amuudlohvp$, and presumably originate
to the long distance control of the correlators with light components.
The $a_{\mu,s/c/\text{disc}}^{\text{{\tiny LO-HVP}}}$ are already determined with high enough precision
and consistent among various LQCD groups.
The window method based on the coordinate space expression
allows a combined analyses using both LQCD and dispersive method and could deliver the estimates of $\amulohvp$
at a per-mil level close upon the target precision.

As a future perspective, the following developments are work in progress;
\textcircled{\scriptsize 1}
significantly improved statistics with continuously increasing ensembles and a better noise control,
\textcircled{\scriptsize 2}
much better understandings of the light connected correlators at large distance and FV effects
via the LQCD data driven estimates for the time-like pion form factor~\cite{Meyer:2018PoS,Erben:2017hvr}
together with the improved bounding method~\cite{Meyer:2018PoS, Lehner:2018PoS},
\textcircled{\scriptsize 3}
precise estimates of SIB/QED corrections
in the stochastic method combined with the perturbative expansion
in $\alpha = e^2_{\rm el}/(4\pi)$ and $\delta m = m_d - m_u$~\cite{Toth:2018PoS}.
The next step required for the LQCD is to establish the consensus among various LQCD estimates of the $\amulohvp$
with a per-mil precision based on the aforementioned progress,
which would be achieved within a few years and have an important indication for
the FNAL-E989, J-PARC-E34, and MUonE experiments with respect to exploring the BSM.

\paragraph{Acknowledgments}
The author thanks to the organisers of the lattice conference 2018 for the invitation to the plenary talk
and the accommodation during the conference. He thanks to GSI
for the financial support for his travel to the conference site (Michigan-USA).
He thanks for fruitful discussions to
H.~Wittig, H.~Meyer, A~.Risch, 
L.~Lellouch, B.~T\'{o}th, 
C.~Davies, R.~Water, 
C.~Lehner, V.~G\"{u}lpers, A.~Meyer, 
S.~Simula, D.~Giusti, 
M.~Marinkovic, T.~Blum, and J.~Zanotti.
\appendix
\section{Summary Table 2018}\label{app:sum_tab}

\begin{table}[!h]
\begin{center}
{\small 
\caption{Summary of SIB and/or QED corrections: $\delta\amulohvp$.}
\label{tab:sib_qed}
\begin{tabular}{l|cc}
\hline\hline
Collab.&\hspace{-2mm}
$10^{-10}\cdot\delta\amulohvp$&\hspace{-3mm}
Comments\\
\hline
ETM-18~\cite{Giusti:2018mdh}&\hspace{-2mm}
7(2)&\hspace{-3mm}
SIB + QED in LO. Only quark-connected.\\
BMW-18~\cite{Borsanyi:2017zdw}&\hspace{-2mm}
7.8(5.1)&\hspace{-3mm}
SIB + QED with XPT/dispersion. \\ 
RBC/UKQCD-18~\cite{Blum:2018mom,Gulpers:2018mim}&\hspace{-2mm}
9.5(10.2)&\hspace{-3mm}
SIB + QED in LO. [conn] + [disc1] in Fig.~\ref{fig:qed}.\\
FHM-18~\cite{Chakraborty:2017tqp}&\hspace{-2mm}
9.5(4.5)&\hspace{-3mm}
Full SIB for ud-conn. Simulation w. $\delta m \neq 0$ but $\alpha = 0$.\\
QCDSF Prelim.~\cite{Zanotti:2018PoS}&\hspace{-2mm}
$\lesssim 1\%$ in total&\hspace{-3mm}
Full QED: Simulation w. $\alpha \neq 0$ but $\delta m = 0$. $m_{\pi}\sim 400$ MeV.\\
\hline\hline
\end{tabular}
} 
\end{center}
\end{table}
\begin{table}[!h]
\begin{center}
\caption{Summary of FV effects:
$\Delta\amuudlohvp(L_1,L_2) = \amuudlohvp(L_2) - \amuudlohvp(L_1)$.
2$\pi$XPT = 2-pion chiral perturbation theory~\protect\cite{Aubin:2015rzx},
GSLM = Gounaris-Sakurai-L\"{u}scher-Meyer method (see Sec.~\protect\ref{subsec:long_distance}),
DQCD = dual-QCD ansatz~\protect\cite{Giusti:2018mdh}.
}\label{tab:fv}
\begin{tabular}{l|c|cc}
\hline\hline
Condition&
Collab.&
Method&
$10^{-10}\cdot \Delta\amuudlohvp$\\
\hline
$(L_1,L_2)=(4.66,6.22)$ fm&
RBC/UKQCD~\cite{Meyer:2018PoS,Lehner:2018PoS}&
2$\pi$XPT&
12.2\\
$m_{\pi} \simeq 135$ MeV.&
{}&
GSLM&
20(3)\\
{}&
{}&
LQCD&
21.6(6.3)\\
\hline
$(L_1,L_2)=(5.4,10.8)$ fm&
PACS Prelim.~\cite{Shintani:2018PoS}&
LQCD&
40(18)\\
$m_{\pi} \simeq 135$ MeV.&
{}&
{}&
{}\\
\hline
$L_2\gg L_1$, $m_{\pi}L_1\sim 4$.&
BMW-18~\cite{Borsanyi:2017zdw}&
2$\pi$XPT&
15(15)\\
{}&
RBC/UKQCD-18~\cite{Blum:2018mom}&
2$\pi$XPT&
16(4)\\
{}&
RBC/UKQCD Prelim.~\cite{Lehner:2018PoS}&
GSLM&
22(1)\\
{}&
Mainz Prelim.~\cite{Gerardin:2018sin}&
GSLM&
20(4)\\ 
{}&
ETM-18~\cite{Giusti:2018mdh}&
GSLM/DQCD&
31(6)\\
\hline\hline
\end{tabular}
\end{center}
\end{table}
\begin{table}[!h]
\begin{center}
\caption{Summary of $\amulohvp$. See also Fig.~\protect\ref{fig:amu_compare}.
In the second column, ``$N_f = $2+1'' takes account of a charm quark in varence quarks but not in sea-quarks.
In $N_f = 2$, a strange quark effect in sea-quarks is also missing.
The first and second brackets show statistical and systematic errors, respectively.
In the case with only one bracket, it includes both errors.
HISQ = highly improved staggered quarks,
Stout4S = 4 steps stout-smeared staggered quarks,
tmQCD = twisted mass QCD,
DWF = domain wall fermions,
Clover = $\mathcal{O}(a)$ improved Wilson quarks.
For the staggered quarks, a rooting technique is used.
TMR = time-momentum representation,
VMD = vector-meson dominance.}
\label{tab:amu_cmp}
\begin{tabular}{l|cccc}
\hline\hline
Collab.&
$N_f$&
$10^{10}\cdot\amulohvp$&
Fermion&
$\hat{\Pi}(Q^2)$\\
\hline
HPQCD-17~\cite{Chakraborty:2016mwy}&
2+1+1&
667(6)(12)&
HISQ&
Pad\'{e} w. Moments\\
FHM-Prelim.~\cite{FHM:private-discuss}&
2+1+1&
690(13)(-)&
HISQ&
Pad\'{e} w. Moments\\
BMW-18~\cite{Borsanyi:2017zdw}&
2+1+1&
711.1(7.5)(17.5)&
Stout4S&
TMR\\
ETM-14~\cite{Burger:2013jya}&
2+1+1&
674(21)(18)&
tmQCD&
VMD\\
ETM-18~\cite{Giusti:2018mdh}&
2+1+1&
683(19)&
tmQCD&
TMR\\
\hline
RBC/UKQCD-18~\cite{Blum:2018mom}&
2+1&
717.4(16.3)(9.2)&
DWF&
TMR\\
Mainz Prelim.~\cite{Gerardin:2018sin}&
2+1&
711($\pm 3\%$)&
Clover&
TMR\\
\hline
Mainz-17~\cite{DellaMorte:2017dyu}&
2&
$654(32)({}^{+21}_{-23})$&
Clover&
TMR\\
\hline
Jegerlehner-18~\cite{Jegerlehner:2018zrj}&
pheno.&
689.46(3.25)&
-&
dispersion\\
DHMZ-17~\cite{Davier:2017zfy}&
pheno.&
693.1(3.4)&
-&
dispersion\\
KNT-18~\cite{Keshavarzi:2018mgv}&
pheno.&
693.37(2.46)&
-&
dispersion\\
\hline
RBC/UKQCD-18~\cite{Blum:2018mom}&
lat.+pheno.&
692.5(1.4)(2.3)&
DWF&
TMR + disp.\\
\hline\hline
\end{tabular}
\end{center}
\end{table}
\begin{table}[!h]
\begin{center}
\caption{$\amulohvp$, flavor by flavor shown in Fig.~\protect\ref{fig:amu_compare_fbyf}.
In the HPQCD result for the light-connected contributions ($\amuudlohvp$, annotated with $\dagger$),
FV and taste-breaking corrections are calculated by using the $\pi - \rho$ effective model where
the corrections associated with {\em disconnected} $\pi-\pi$ diagrams has not been excluded.
If this correction is excluded, the results would become somewhat larger.
For details, consult with the authors of Ref.~\protect\cite{Chakraborty:2016mwy}.}
\label{tab:amu_fbf}
{\small
\begin{tabular}{l|ccccc}
\hline\hline
Collab.&\hspace{-2mm}
$N_f$&\hspace{-3mm}
$\amuudlohvp$&\hspace{-3mm}
$\amuslohvp$&\hspace{-3mm}
$\amuclohvp$&\hspace{-3mm}
$\amudisclohvp$\\
\hline
HPQCD-17/14~\cite{Chakraborty:2016mwy}&\hspace{-2mm}
2+1+1&\hspace{-3mm}
$599.0(6.0)(11.0)^{\dagger}$&\hspace{-3mm}
53.41(00)(59)&\hspace{-3mm}
14.42(00)(39)&\hspace{-3mm}
0(9)(-)\\
FHM-Prelim.~\cite{FHM:private-discuss}&\hspace{-2mm}
2+1+1&\hspace{-3mm}
630(8)($-$)&\hspace{-3mm}
$-$&\hspace{-3mm}
$-$&\hspace{-3mm}
$-$\\
BMW-18~\cite{Borsanyi:2017zdw}&\hspace{-2mm}
2+1+1&\hspace{-3mm}
647.6(7.5)(17.7)&\hspace{-3mm}
53.73(0.04)(0.49)&\hspace{-3mm}
14.74(0.04)(0.16)&\hspace{-3mm}
-12.8(1.1)(1.6)\\
ETM-18/17~\cite{Giusti:2018mdh}&\hspace{-2mm}
2+1+1&\hspace{-3mm}
619.0(17.8)&\hspace{-3mm}
53.1(1.6)(2.0)&\hspace{-3mm}
14.75(42)(37)&\hspace{-3mm}
$-$\\
\hline
RBC/UKQCD-18~\cite{Blum:2018mom}&\hspace{-2mm}
2+1&\hspace{-3mm}
649.7(14.2)(4.9)&\hspace{-3mm}
53.2(4)(3)&\hspace{-3mm}
14.3(0)(7)&\hspace{-3mm}
-11.2(3.3)(2.3)\\
Mainz Prelim.~\cite{Gerardin:2018sin}&\hspace{-2mm}
2+1&\hspace{-3mm}
643(21.0)(-)&\hspace{-3mm}
54.0(2.2)(0.8)&\hspace{-3mm}
14.95(0.47)(0.11)&\hspace{-3mm}
$-$\\
\hline
Mainz-17~\cite{DellaMorte:2017dyu}&\hspace{-2mm}
2&\hspace{-3mm}
588.2(31.7)(16.6)&\hspace{-3mm}
51.1(1.7)(0.4)&\hspace{-3mm}
14.3(2)(1)&\hspace{-3mm}
$-$\\
\hline\hline
\end{tabular}
} 
\end{center}
\end{table}
\begin{table}
\begin{center}
\caption{$\Pi_{1,2}$: up/down-contributions and total.
The light component $\Pi_{n}^{\rm ud}$ includes FV corrections.
The total results, $\Pi_{n}^{\rm tot},$ take account SIB/QED corrections also.
For HPQCD-17 (annotated with $\dagger$), we have multiplied a charge factor $(9/5)$
to the original numbers (plus corrections) in Refs.~\protect\cite{Chakraborty:2016mwy}
to make them being a same convention as those in the other groups.}
\label{tab:tm_fbf}
{\small
\begin{tabular}{l|ccccc}
\hline\hline
Collab.&\hspace{-2mm}
$N_f$&\hspace{-3mm}
$\Pi_1^{\rm ud}$&\hspace{-3mm}
$-\Pi_2^{\rm ud}$&\hspace{-3mm}
$\Pi_1^{\rm tot}$&\hspace{-3mm}
$-\Pi_2^{\rm tot}$\\
\hline
HPQCD-17~\cite{Chakraborty:2016mwy}&\hspace{-2mm}
2+1+1&\hspace{-3mm}
${0.1606(22)(14)}^{\dagger}$&\hspace{-3mm}
${0.362(7)(14)}^{\dagger}$&\hspace{-3mm}
0.0984(14)&\hspace{-3mm}
0.2070(89)\\
BMW-17~\cite{Borsanyi:2016lpl}&\hspace{-2mm}
2+1+1&\hspace{-3mm}
0.1665(17)(52)&\hspace{-3mm}
0.327(10)(23)&\hspace{-3mm}
0.1000(10)(28)&\hspace{-3mm}
0.181(6)(11)\\
ETM-18~\cite{Giusti:2018mdh}&\hspace{-2mm}
2+1+1&\hspace{-3mm}
0.1642(33)&\hspace{-3mm}
0.383(16)&\hspace{-3mm}
$-$&\hspace{-3mm}
$-$\\
\hline
RBC/UKQCD-18~\cite{Blum:2018mom}&\hspace{-2mm}
2+1&\hspace{-3mm}
0.1713(46)(14)&\hspace{-3mm}
0.352(37)(10)&\hspace{-3mm}
$-$&\hspace{-3mm}
$-$\\%
\hline
Benayoun-16~\cite{Benayoun:2016krn}&\hspace{-2mm}
pheno.&\hspace{-3mm}
$-$&\hspace{-3mm}
$-$&\hspace{-3mm}
0.09896(73)&\hspace{-3mm}
0.20569(162)\\
Charles-18~\cite{Charles:2017snx}&\hspace{-2mm}
pheno.&\hspace{-3mm}
$-$&\hspace{-3mm}
$-$&\hspace{-3mm}
0.10043(36)&\hspace{-3mm}
0.20914(113)\\
\hline\hline
\end{tabular}
} 
\end{center}
\end{table}


\end{document}